\documentclass[lettersize,journal]{IEEEtran}
\usepackage{amsmath,amssymb,amsfonts}
\usepackage{algorithmic}
\usepackage{algorithm}
\usepackage{array}
\usepackage{textcomp}
\usepackage{stfloats}
\usepackage{url}
\usepackage{verbatim}
\usepackage{graphicx}
\usepackage{bbm}
\usepackage{epstopdf}
\usepackage[numbers,sort&compress]{natbib}
\newtheorem{definition}{Definition}
\newtheorem{assumption}{Assumption}
\newtheorem{lemma}{Lemma}
\newtheorem{theorem}{Theorem}
\newtheorem{problem}{Problem}
\newtheorem{remark}{Remark}
\usepackage{tikz}
\usepackage{mathtools}
\usepackage{cuted}

\DeclareMathOperator{\atantwo}{atan2}
\usepackage{printlen}
\usepackage[labelformat=simple]{subcaption}

\usepackage{array, xcolor, colortbl, caption}
\begin{document}

\title{Target Enclosing Control for Nonholonomic Multi-Agent Systems  with Connectivity Maintenance and Collision Avoidance}
\author{Boyin Zheng, Yahui Hao, Lu Liu, \IEEEmembership{Senior Member,~IEEE}
	
\thanks{The authors are with the Department of Mechanical Engineering, 	City University of Hong Kong, Hong Kong, China (e-mail:boyizheng2-c@my.cityu.edu.hk; yahuihao2-c@my.cityu.edu.hk; luliu45@cityu.edu.hk). (Corresponding author: Lu Liu.)}}




\maketitle

\begin{abstract}
    This article addresses the moving target enclosing control problem for nonholonomic multi-agent systems with guaranteed network connectivity and collision avoidance.  We propose a novel control scheme to handle distance constraints imposed by the agents' limited interaction ranges and collision-free thresholds. By leveraging a Henneberg construction method, we innovatively formulate the target enclosing requirements within an isostatic distance-based formation framework, facilitating the integration of distance constraints.   Compared with existing results, our  approach ensures the positive definiteness of the underlying rigidity matrix and does not require controlling the target’s motion. 
   To eliminate the occurrences of control singularities caused by nonholonomic constraints, we propose a fixed-time angular control law using barrier Lyapunov functions.   Additionally, we develop a linear velocity control law using the prescribed performance control approach and transformed error constraints. 
  We rigorously prove that our control laws enable the multi-agent system to asymptotically achieve the desired angular formation pattern around a moving target while satisfying the established distance constraints. Finally, a simulation example is provided to validate the effectiveness of the proposed method.
\end{abstract}


\section{Introduction}
Target enclosing control problems have garnered significant attention in recent years, owing to their extensive applications in critical areas such as support and escort operations \cite{yu2016coordinated}, spatial mapping and data acquisition \cite{li2020cooperative}, 
and environmental protection \cite{wang2023oa,Bono2022target}. In these problems, a group of mobile agents is tasked with surrounding a target--stationary or moving--while achieving some prescribed angular formation pattern. Early studies in this area focused on static target enclosing with single-integrator-type multi-agent systems (MASs) \cite{kim2007cooperative,wang2013forming,wang2018formation}.
In particular, the consensus-based control laws proposed in \cite{wang2013forming} can achieve unevenly spaced formation patterns around a circle and were later extended to handle MASs with more complex dynamics, such as double integrators \cite{wang2023coverage}, Euler–Lagrange systems \cite{ma2023cooperative} and nonholonomic dynamics \cite{yu2018circular}.

Recently, more attention has been paid to moving target enclosing control for MASs. 
In \cite{dou2020target,dou2021moving}, distributed control protocols combined with target state estimators were proposed to handle uncertainties in the target’s velocity for integrator-type MASs. Beyond agents with integrator-type dynamics, plenty of effort has also been devoted to addressing nonholonomic MASs \cite{zheng2015enclosing,yu2017cooperative, yu2018cooperative,peng2021cooperative,sen2021circumnavigation,Milad2020flocking}. For example, \cite{yu2017cooperative} developed a backstepping approach with the stabilization analysis conducted using reduction theorems, yet it only achieved evenly spaced formations. Based on the linearization method, a prescribed unevenly spaced formation was achieved in \cite{peng2021cooperative}. The study in \cite{sen2021circumnavigation} further tackled more complex scenarios in target formation control, including input constraints, non-identical circling radii, and incomplete target information. 
\textcolor{black}{In \cite{JU2023mpc}, input constraints and unknown external disturbances were studied simultaneously  for nonholonomic MASs with model predictive control strategies. In addition to these efforts, the target enclosing problem with unknown target velocities was investigated in \cite{ju2023enclosing, zheng2025robust} using  adaptive and robust control techniques.}


In practical scenarios, ensuring network connectivity and collision avoidance is crucial to achieving the objectives of cooperative control tasks and guaranteeing the safe deployment of MASs.  To handle distance constraints due to these safety concerns, artificial potential function-based methods have been widely adopted. However, they often encounter local minima issues \cite{fu2022distributed}. To overcome this challenge, a navigation function-based approach was proposed for formation stabilization tasks in first-order MASs \cite{loizou2021connectivity}. An alternative strategy to avoid local minima issues involves employing a transformation technique within the potential function design \cite{zheng2023cyclic}. Instead of directly utilizing distance constraints, transformed error constraints were used in the design of the potential function whose value reached its minimum only when the desired formation was achieved. 
Furthermore, the convergence performance of MASs, including transient responses, convergence rates, and steady-state errors, is also a critical consideration in some cooperative control tasks. 
Prescribed performance control (PPC), initially introduced in \cite{Bechlioulis2008robust}, enables the system to achieve desired transient responses and arbitrarily small steady-state errors. Over recent years, PPC protocols have been widely applied to formation control problems \cite{chen2020leader,mehdifar2020prescribed,ke2022fixed,xiong2022cooperative}. 
In particular, \cite{mehdifar2020prescribed} and \cite{ke2022fixed} investigated constrained distance-based formation stabilization and maneuvering problems, respectively. In the context of \textcolor{black}{the}  target enclosing control problem, \cite{xiong2022cooperative} and \cite{lu2023cooperative} adopted  a feedback linearization method  for nonholonomic vehicles and leveraged PPC methods  to handle  distance constraints. However, only \textcolor{black}{an} evenly-spaced formation pattern was achieved in \cite{xiong2022cooperative} and both works did not consider  connectivity issue\textcolor{black}{s} between the agents.

Motivated by the observations above, this paper investigates the target enclosing problem for unicycle-type MASs, with a focus on connectivity maintenance and collision avoidance. This problem entails three primary challenges. 
	{\it First}, existing results for nonholonomic vehicles in this field rely on two types of approaches. The first approach employs a virtual circular motion center, derived from each agent's actual position, to track the target \cite{zheng2015enclosing,yu2018cooperative,yu2017cooperative}. Due to its complexity, few studies have integrated distance constraints into this method. The second approach is the feedback linearization method \cite{lu2023cooperative, peng2021cooperative}. Although this design is more straightforward \textcolor{black}{in incorporating} distance constraints and its convergence analysis is simplified, it may introduce sensitivity to model uncertainties and only \textcolor{black}{achieve} desired configurations at the offset points, rather than at the centers of the MAS. 
{\it Second}, although distance-based formation frameworks have the intrinsic advantage of incorporating distance constraints,  transforming the desired circling radius and the prescribed angular formation pattern of a target enclosing problem into an isostatic formation framework remains to be addressed. The isostatic property of a framework, linked to the positive definiteness of the underlying rigidity matrix, is essential for the convergence analysis of the closed-loop system.
{\it Third}, the boundedness and convergence of the closed-loop system for constrained distance-based formation control present significant analytical difficulties. These difficulties arise due to the utilization of highly nonlinear controllers and the unicycles' nonholonomic constraints, which are considered in this work. 
The key contributions of this research can thus be summarized as follows.
\begin{itemize}
	\item We propose an effective control scheme under which the MAS can achieve an unevenly spaced formation pattern around a moving target without violating the distance constraints imposed by the agents' limited interaction ranges and collision-free thresholds.  Unlike previous results \cite{zheng2015enclosing,yu2017cooperative, yu2018cooperative,peng2021cooperative,sen2021circumnavigation,Milad2020flocking,xiong2022cooperative,lu2023cooperative}, which either did not consider distance constraints or addressed them inadequately, our approach fully integrates network connectivity and collision avoidance among all agents and the target. Compared to our previous result \cite{zheng2023cyclic}, this work incorporates a PPC approach and a universal barrier Lyapunov function, which allows for asymmetric error bounds. These improvements not only relax the assumptions on the initial states of the MAS but also guarantee the prescribed transient and steady-state responses of the system.
	\item  We transform target enclosing requirements into an isostatic distance-based formation framework by performing  \textcolor{black}{an incremental} procedure of \textcolor{black}{Henneberg} construction. Furthermore, in existing results on constrained distance-based formation control \cite{mehdifar2020prescribed,ke2022fixed},  the number of vertices on the formation framework needs to match the number of controllable agents.  In contrast, the formation framework constructed in our work includes the target as one of its vertices but does not require controlling its motion. This characteristic aligns more closely with real-world applications. 
	In particular, we reduce the order of the original rigidity matrix by a decomposition operation, yielding a sub-matrix that preserves some essential properties for subsequent convergence analysis.
	\item We conduct a rigorous theoretical analysis demonstrating the boundedness and convergence of the closed-loop system under the proposed laws. This analysis is achieved by unifying rigidity graph theory, barrier Lyapunov theory, and input-to-state stability theory. Moreover, we introduce a fixed-time angular control law derived from a barrier Lyapunov function. This control law not only eliminates the occurrences of control singularities but also ensures fixed-time convergence of the heading angle error to zero. As a result, it simplifies the formation convergence analysis by removing some auxiliary terms.
\end{itemize}

\textit{Notations:} $\mathbb{R}^+$ represents all positive real numbers.  $\mathbb{R}^n$ and $\mathbb{R}^{m\times n}$ denote the space of all real $n$-dimensional vectors and the space of all real $m\times n$ matrices, respectively. For any vector $x \in \mathbb{R}^n$, $||x||$ denotes its Euclidean norm. $\boldsymbol{0}_n$ and $\boldsymbol{1}_n \in \mathbb{R}^n$ represent vectors with all their elements being $0$ and $1$, respectively. $I_n \in \mathbb{R}^{n\times n}$ is the identity matrix. For a matrix $P\in \mathbb{R}^{m\times n}$, $P^T$ denotes its transpose. The Kronecker product of matrices $P \in \mathbb{R}^{m\times n}$ and $Q \in \mathbb{R}^{p\times q}$ is denoted as $P \otimes Q$. For any positive integer $N$, define $N^-=N-1$ and $N^+=N+1$.
Given a graph $\mathcal{G}^*=\{\mathcal{V}^*, \mathcal{E}_{ord}\}$ where $\mathcal{V}^*=\{0,1,\dots,N\}$ defines the vertex set and $\mathcal{E}_{ord} =\{(1,2),(2,3),\ldots,(N^-,N),(1,0),(2,0),\ldots,(N,0)\}$ denotes an ordered edge set. Let a set of real variables be $x_{ij}\coloneqq x_{(i,j)} \in \mathbb{R}$ where $ (i,j) \in \mathcal{E}_{ord}$, then $\mathrm{col}(x_{ij})\coloneqq [x_{12},x_{23},\ldots,x_{N^-N},x_{10},x_{20},\ldots,x_{N0}]^T \in \mathrm{R}^{2N-1}$ defines a column vector and $\displaystyle \mathrm{diag}(x_{ij})\in \mathrm{R}^{(2N-1)\times(2N-1)}$ defines a diagonal matrix with diagonal elements being $x_{12},x_{23},\ldots,x_{N^-N},x_{10},x_{20},\ldots,x_{N0}$, respectively.
For any number $\mu\in \mathbb{R}^+$ and $x\in \mathbb{R}$, the function $\lceil x\rfloor^\mu \coloneqq |x|^\mu \mathrm{sign}(x)$.
Define the rotational matrix with rotation angle $\delta\in [-\pi,\pi]$ by $R_{\delta}=\begin{pmatrix}\cos\delta & -\sin\delta \\ \sin\delta & \cos\delta \end{pmatrix}$. \color{black} Table  \ref{t1} summarizes the key symbols and their explanations in this article.

\begin{table}[ht]
	\caption{Key symbols and the explanations.}
	\label{t1}
	\renewcommand{\arraystretch}{1.4}
   \arrayrulecolor{black}
	\begin{tabular}{|>{\color{black}\centering\arraybackslash}m{2.8cm}|>{\color{black}\arraybackslash}m{5.0cm}|}
		\hline \textbf{Symbol} & \textbf{Explanation} \\
		\hline $e_{ij}(t)$  
		 & distance formation error of edge $(i,j)\in \overline{\mathcal{E}} $ \\
		\hline $\underline{e}_{ij}^*, \overline{e}_{ij}^*$ 
		 & CM$\&$CF and {\it IR}  induced error constraints  such that
		 $-\underline{e}_{ij}^*<e_{ij}(t)<\overline{e}_{ij}^*$ \\
		\hline $\eta_{ij}(t)$  
		& squared distance formation error of edge $(i,j)\in \overline{\mathcal{E}}$\\
		\hline $\underline{\eta}_{ij}, \overline{\eta}_{ij}$ 
		& squared error constraints such that 
		$	-\underline{\eta}_{ij}<\eta_{ij}(t)<\overline{\eta}_{ij}$ \\
		\hline $	\beta_{ij}(t)$ 
		& prescribed performance function of edge $(i,j)\in \overline{\mathcal{E}} $  \\
		\hline  $\underline{\eta}_{ij}(t),\overline{\eta}_{ij}(t)$ 
		& positive decreasing functions  such that 
		  $-\underline{\eta}_{ij}(t)<\eta_{ij}(t)<\overline{\eta}_{ij}(t)$\\
		\hline $\xi_{ij}(t)=\frac{{\eta}_{ij}(t)}{\beta_{ij}(t)}$  & internal error variable of edge $(i,j)\in \overline{\mathcal{E}} $ \\
		\hline $\underline{\xi}_{ij}, \overline{\xi}_{ij}$ 
		& constant internal error bounds such that $	-\underline{\xi}_{ij}<\xi_{ij}(t)<\overline{\xi}_{ij}$ \\
		\hline $\sigma_{ij}(t)$  
		 & transformed error function of edge $(i,j)\in \overline{\mathcal{E}} $ \\
		\hline $e_{\theta_i}(t)$ 
		& heading angle error of agent $i\in {\mathcal{V}} $ \\
		\hline $\sigma_{{e}_{\theta_i}}(t)$  
		& transformed angular error function of agent $i\in {\mathcal{V}} $ \\
		\hline
	\end{tabular}
\end{table}

\color{black}
This article is structured as follows. The preliminaries and problem formulation are provided in Section \ref{II} and Section \ref{III}, respectively. The distributed target enclosing control laws are developed in Section \ref{IV}. In Section \ref{V}, the convergence analysis of the closed-loop system is presented. Section \ref{VI} illustrates a simulation example. Finally, Section \ref{VII} concludes the article.

\section{Preliminaries}\label{II}
\subsection{Rigid Graph Theory}\label{rigidgraphtheory}
The underlying sensing graph of the MAS is defined by an undirected graph $\mathcal{G}=\{\mathcal{V}, \mathcal{E}\}$ where $\mathcal{V}=\{1,\ldots,N\}$ denotes the set of agents in the MAS and $\mathcal{E}=\{(i,i+1)\colon i\in \mathcal{V}\setminus \{N\}\}$ is the edge set. The neighbor set of agent $i$ is defined as $\mathcal{N}_i=\{j\in \mathcal{V} \colon (i,j)\in \mathcal{E}\}$. Denote the target vertex by $0$, the overall sensing graph among the target and the MAS is defined by $\overline{\mathcal{G}}=\{\overline{\mathcal{V}}, \overline{\mathcal{E}}\}$ where $\overline{\mathcal{V}}=\{0,\mathcal{V}\}$ is the vertex set and $\overline{\mathcal{E}}=\{(i,0)\colon i \in \mathcal{V}\}\cup \mathcal{E}$ represents the edge set. Under the overall sensing topology, the neighbor set of agent $i \in \mathcal{V}$ is redefined as $\overline{\mathcal{N}}_i=\{0\}\cup \mathcal{N}_i$. Note that the overall sensing graph contains $N+1$ vertices and $2N-1$ edges. 

Define the coordinates of the vertices in graph $\overline{\mathcal{G}}$ as $\overline{p}=[p_1^T,\ldots,p_N^T,p_0^T]^T \in \mathbb{R}^{2N+2}$, then a formation framework $\mathcal{F}=\left(\overline{\mathcal{G}},\overline{p}\right)$ can be obtained. For simplicity of the following analysis, an edge function $\phi\left(\overline{p}\right)\colon \mathbb{R}^{2N+2}\mapsto\mathbb{R}^{2N-1}$ of a given ordered edge set $\{(1,2),(2,3),\ldots,(N^-,N),(1,0),(2,0),\ldots,(N,0)\}$ is defined as 
{\small\begin{equation}
		\begin{aligned}
			\phi\left(\overline{p}\right)=\bigg[\|p_1-p_2\|^2,\|p_2-p_3\|^2,\ldots, \|p_{N^-}-p_N\|^2,\\
			\|p_1-p_0\|^2, \|p_2-p_0\|^2,\ldots, \|p_N-p_0\|^2\bigg]^T.\\
		\end{aligned}
\end{equation}}
\noindent Now the rigidity matrix $R\left(\overline{p}\right)\coloneqq\frac{1}{2}\frac{\partial\phi}{\partial\overline{p}} \in \mathbb{R}^{(2N-1)\times(2N+2)}$ of framework $\mathcal{F}$ can be expressed as 
{\small \begin{equation}\label{RigidityMatrix}
		\begin{bmatrix}
			p_{12}^T&-p_{12}^T&0&\cdots&0&0&0\\
			0&p_{23}^T&-p_{23}^T&\cdots&0&0&0\\
			\vdots&\vdots&\vdots&\vdots&\vdots&\vdots&\vdots\\
			0&0&0&\cdots&p_{N^-N}^T&-p_{N^-N}^T&0\\
			p_{10}^T&0&0&\cdots&0&0&-p_{10}^T\\
			0&p_{20}^T&0&\cdots&0&0&-p_{20}^T\\
			\vdots&\vdots&\vdots&\vdots&\vdots&\vdots&\vdots\\
			0&0&0&\cdots&0&p_{N0}^T&-p_{N0}^T\\
		\end{bmatrix},
\end{equation}}

\noindent where $p_{ij}=p_i-p_j$ represents the relative position between agent $i$ and its neighbor $j$ in the overall sensing graph. In the rigid graph theory, a rigid framework implies that by preserving the distances of a given edge set $\overline{\mathcal{E}}$, one can guarantee that all distances among the vertices of the graph remain unchanged. In other words, the shape of the formation is preserved. It is known that in $\mathbb{R}^2$, $\mathrm{rank}\left(R\left(\overline{p}\right)\right)\leq 2\left|\overline{\mathcal{V}}\right|-3$ \cite{asimow1979rigidity}. A framework is minimally rigid if it is rigid and the removal of a single edge causes it to lose rigidity. In $\mathbb{R}^2$, a rigid framework is minimally rigid if and only if $\left|\overline{\mathcal{E}}\right|=2\left|\overline{\mathcal{V}}\right|-3$ \cite{anderson2008rigid}. A framework is infinitesimally rigid if and only if every infinitesimal motion is an Euclidean motion, which is generated by rotations and translations in $\mathbb{R}^2$. Moreover, a rigid framework is infinitesimally rigid if and only if $\mathrm{rank}\left(R\left(\overline{p}\right)\right)=2\left|\overline{\mathcal{V}}\right|-3$ \cite{de2019formation}. Finally, a {\it minimally and infinitesimally rigid} ({\it MIR}) framework is also said to be isostatic. 

\color{black}
\subsection{Henneberg Construction in the Plane   }\label{HC}

	A \textit{Henneberg construction} is an inductive method for constructing minimally rigid graphs, also known as Laman graphs in the plane \cite{tay1985generating}. Starting from a base graph--typically a single edge connecting two vertices--the construction proceeds by repeatedly applying one of the following two operations:
	
	\begin{itemize}
		\item \textbf{Henneberg I }(Vertex addition): add a new vertex and connect it to two existing vertices with new edges.
		\item \textbf{Henneberg II }(Edge splitting): remove an existing edge between two vertices, add a new vertex, and connect this new vertex to the two vertices previously connected by the removed edge and to a third existing vertex.
	\end{itemize}
	
 Through conducting these two operations successively, we can ensure that the resulting graph is minimally rigid in two dimensions. 

\color{black}
\subsection{Useful Lemmas}\label{lemmas}
Next, some useful lemmas regarding properties of the rigidity matrix, fixed-time stability and the Picard-Lindelöf theorem for an initial value problem will be introduced.

\begin{lemma}\label{translation invariance}
	\cite{de2019formation} For any vector $v\in\mathbb{R}^2$, it follows that $R\left(\overline{p}\right)\left(1_{N+1}\otimes v\right)\ =\boldsymbol{0}_{2N-1}$. 
\end{lemma}
\begin{lemma}\label{Rigidity properties}
	\cite{de2019formation} If a framework is {\it MIR} or isostatic in $\mathbb{R}^2$, then its rigidity matrix has full row rank and $R\left(\overline{p}\right)R\left(\overline{p}\right)^T$ is positive definite.
\end{lemma}
\begin{lemma}\label{Transformederrorconstraints}
	\cite{cai2015formation} If $\mathcal{F}^\ast=(\overline{\mathcal{G}},\overline{p}^*)$ is infinitesimally rigid, then there exists a small positive constant $\nu$ such that all frameworks $\mathcal{F}=(\overline{\mathcal{G}},\overline{p})$ that satisfy $\Psi(\mathcal{F}^\ast, \mathcal{F})\coloneqq\sum_{(i,j)\in\mathcal{\overline{E}}}(\|p_i^*-p_j^*\|-\|p_i-p_j\|)^2\leq \nu$, are also infinitesimally rigid.
\end{lemma}
\begin{lemma}\label{fixedtime}
	\cite{Polyakov2012nonlinear} If there exists a continuous radially unbounded and positive-definite function $V\left(x\right)$ such that
	\begin{equation}
		\dot{V}(x)\leq -\alpha V^p(x)-\rho V^q(x),
	\end{equation}
	for constants $\alpha>0$, $\rho>0$, $p>1$ and $0<q<1$, then the origin of the system is globally fixed-time stable and the settling time function $T$ can be estimated by
	\begin{equation}
		T\leq T_{max} \coloneqq \frac{1}{\alpha(p-1)}+\frac{1}{\rho(1-q)}.
	\end{equation}
\end{lemma}
\begin{lemma}\label{IVPsolutiuon} 
	\cite{sontag2013mathematical} Given an initial-value problem (IVP)
	\begin{equation}
		\dot{x}=f(t,x(t)), \quad  x(0) \in \mathcal{X},
	\end{equation}
	assume that $f(t,x): \mathcal{I}\times\mathcal{X} \mapsto \mathbb{R}^N$, where $\mathcal{I}$ is an interval in $\mathbb{R}$ and $\mathcal{X}$ is a subset of $\mathbb{R}^N$, satisfies 1) $f(\cdot,x):\mathcal{I} \mapsto \mathbb{R}^N$ is measurable for each fixed $x$; and 2) $f(t,\cdot): \mathbb{R}^N \mapsto \mathbb{R}^N$ is continuous for each fixed $t$. If $f(t,x)$ is continuous in $t$ and locally Lipschitz in $x$, then for each pair initial condition $(0,x(0))\in \mathcal{I}\times\mathcal{X}$, there exists a unique local solution $\overline{x}(t) \in \mathcal{X}$, also called the maximal solution of the IVP
	on some subinterval $\mathcal{J} \subseteq \mathcal{I}$.
\end{lemma}

\section{Problem Formulation}\label{III}
Consider a unicycle-type MAS moving freely on a two-dimensional plane where the position and the heading of each unicycle are denoted by $p_i$ and $\theta_i$, respectively. Suppose that the dynamics of each agent is described by
\begin{equation}\label{dyna}
	\left[\begin{matrix}{\dot{p}}_i\\{\dot{\theta}}_i\\\end{matrix}\right]=\left[\begin{matrix}{\dot{x}}_i\\{\dot{y}}_i\\{\dot{\theta}}_i\\\end{matrix}\right]=\left[\begin{matrix}v_i\cos\theta_i\\v_i\sin\theta_i\\w_i\\\end{matrix}\right], i=1,\dots,N,
\end{equation}
where $v_i, w_i \in \mathbb{R} $ represent the linear velocity and the angular velocity control inputs, respectively. Define the compact form of the positions and velocities for the MAS by $p=[p_1^T,\ldots,p_N^T]^T \in \mathbb{R}^{2N}$ and $\dot{p}=[\dot{p}_1^T,\ldots,\dot{p}_N^T]^T \in \mathbb{R}^{2N}$, respectively.
The dynamics of the target is described by the following single-integrator model,
\begin{equation}\label{target}
	\dot{p}_0=v_0,
\end{equation}
where $p_0\in \mathbb{R}^2$ corresponds to its current location and $v_0\in \mathbb{R}^2$ represents its velocity which is upper bounded by $\overline{v}_0 \in \mathbb{R^+}$, that is, $\|v_0\|<\overline{v}_0$. 
In the target enclosing control problem, the MAS converges to a circle around the target with a desired radius $r$ and prescribed inter-agent separation angles defined by a set $\boldsymbol{c}=[c_{12}, c_{23}, \ldots, c_{N^-N},c_{N1}]$, where each element $c_{ii^+}\in (0^\circ,180^\circ)$ denotes the desired separation angle between the $i$-th agent and the $(i+1)$-th agent. It is noted that the sum of all angles equals $2\pi$ and that the last element $c_{N1}$ can be uniquely decided given all the other elements of set $\boldsymbol{c}$. 
\subsection{Transforming Target Enclosing Requirements into Distance-based Formation Framework}
\label{Construction of distance-based formation}
\begin{figure}[htbp]
	\centering
	\includegraphics[width=\linewidth]{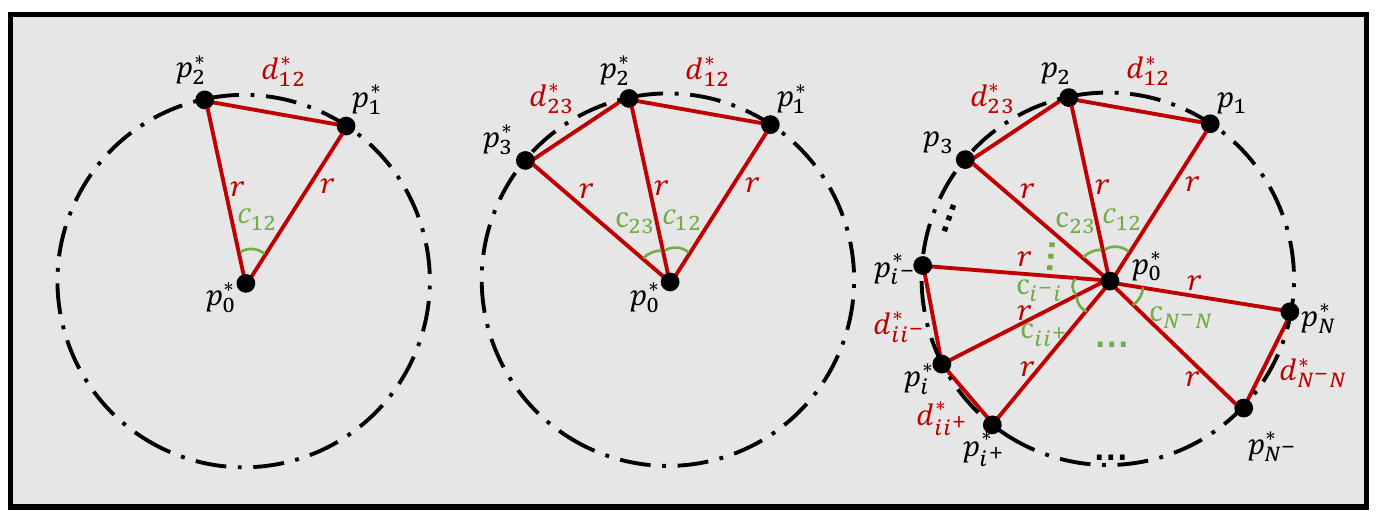}
	\vspace{-6pt}
	\captionsetup{justification=centering}
	\caption{Notations and generation of the desired formation framework.} 
\label{task}
\end{figure}

To facilitate the incorporation of distance constraints induced by connectivity maintenance and collision avoidance problems, these prescribed inter-agent angular distances need to be transformed into some properly chosen desired distances in Euclidean space. This allows us to recast the enclosing objective into a distance-based formation problem suitable for control design. 
Inspired by Section \ref{HC}, we propose the following \textcolor{black}{incremental} procedure of \textcolor{black}{Henneberg} construction tailored to our target enclosing formation problem:

\textit{1) Incremental Henneberg I-Based Construction.} To generate a minimally rigid graph incorporating the desired inter-agent angular distances, we apply an incremental procedure inspired by the Henneberg construction. This method constructs a geometric formation in terms of desired inter-agent and agent-target Euclidean distances.
Specifically, we propose the following procedure:
\begin{itemize}
	\item[i.] construct a triangle with vertices $p_0^*$, $p_1^*$ and $p_2^*$ such that the desired edge lengths satisfy $\|p_{12}^*\|=d_{12}^*\coloneqq 2r\sin{\frac{c_{12}}{2}}$, $\|p_{10}^*\|=d_{10}^*\coloneqq r$, and $ \|p_{20}^*\|=d_{20}^*\coloneqq r$; 
    \item[ii.] add a new vertex $p_i$ and connect it to the target $p_0^*$ and to its previous agent $p_{i^-}^*$ with edge lengths satisfying $\|p_{i0}^*\|=d^*_{i0}\coloneqq r$, $\|p_{i^-i}^*\|=d_{i^-i}^*\coloneqq 2r\sin{\frac{c_{i^-i}}{2}}$;
    \item[iii.] repeat the second step in an incremental manner for $i=3,\ldots, N$.
\end{itemize}
Fig.~\ref{task} illustrates this process, where the black solid circles represent the agents and the red line segments correspond to the desired distances of framework $\mathcal{F}^\ast$. From left to right, the three subfigures show: (i) the initial triangle, (ii) a single vertex addition, and (iii) the completed fan-like structure with all agents positioned incrementally on a circle of radius $r$ centered at the target $p_0^*$.    This setup ensures that each agent respects both its radial distance to the target and the angular spacing to its neighbors.

\textit{2) Guaranteeing Isostatic Formations.} Although the Henneberg construction guarantees that the resulting graph is minimally rigid, it does not necessarily imply that the realized framework is infinitesimally rigid. 
In our specific application, we require the constructed formation framework to be isostatic to effectively apply properties of the rigidity matrix (as discussed in Lemma~\ref{Rigidity properties}). To this end, we invoke the following result from rigidity theory:

\begin{lemma}[\cite{whiteley1988infinitesimally}]\label{isostatic}
	In $\mathbb{R}^2$, if a framework is isostatic, then adding or removing a 2-valent vertex, with the edges not collinear, results in a new plane-isostatic framework.
\end{lemma}

Based on this result, we can now verify that the formation generated by our proposed construction procedure satisfies the isostatic property required for our control framework.

\begin{lemma}\label{isostatic_proof}
	The formation framework $\mathcal{F}^\ast=(\overline{\mathcal{G}},\overline{p}^*)$ is {\it MIR} or isostatic.
\end{lemma}

\begin{IEEEproof}
	Denote the triangle constructed from the first three vertices $p_0^*$, $p_1^*$, and $p_2^*$ by $\mathcal{F}_1$. This initial configuration exhibits \textcolor{black}{isostatic} characteristics. At each subsequent vertex addition, the new agent is connected to the target and its predecessor based on the prescribed inter-agent angular pattern $\boldsymbol{c} = [c_{12}, c_{23}, \ldots, c_{N^-N}, c_{N1}]$. Since each angle $c_{ii^+}$ lies in the open interval $(0^\circ, 180^\circ)$, the two connecting edges are guaranteed to be non-collinear. It then follows from Lemma~\ref{isostatic} and mathematical induction that the resulting formation framework $\mathcal{F}^\ast=(\overline{\mathcal{G}}, \overline{p}^*)$ is also isostatic.
\end{IEEEproof}

Denote the actual distance from agent $i$ to its neighbor $j$ by $d_{ij}(t)\coloneqq \|p_{ij}(t)\|$, $(i,j)\in \overline{\mathcal{E}}$. Then the formation errors for the $2N-1$ edges in $\overline{\mathcal{E}}$ are defined as follows,
\begin{eqnarray}\label{formationerrors}
\left\{
\begin{aligned}[c]
	e_{ii^+}(t)\coloneqq d_{ii^+}(t)-d_{ii^+}^*, & \quad i\in \mathcal{V}\setminus\{N\},\\
	e_{i0}(t)\coloneqq d_{i0}(t)-r, \quad\quad&\quad i\in\mathcal{V}.
\end{aligned}
\right.
\end{eqnarray}

\subsection{Connectivity Maintenance and Collision-Free (CM$\&$CF) Induced Error Constraints}\label{ErrorConstraints}

Suppose that neighboring agents $(i,j)\in \overline{\mathcal{E}}$ have limited interaction ranges and collision-free operating thresholds, denoted by
$\overline{d}_{ij}$ and $\underline{d}_{ij}$, respectively. Connectivity maintenance and collision avoidance can be guaranteed only when the actual distance between neighbors is always kept within the set $(\underline{d}_{ij}, \overline{d}_{ij})$, that is, $\underline{d}_{ij}<d_{ij}(t)<\overline{d}_{ij}$, for all $t\geq 0$. 
Alternatively, these distance constraints can be expressed in terms of formation error constraints as follows,
\begin{eqnarray}
\left\{
\begin{aligned}[c]
	-\underline{e}_{ii^+}<e_{ii^+}(t)<\overline{e}_{ii^+}, & \quad i\in \mathcal{V}\setminus\{N\},\\
	-\underline{e}_{i0}<e_{i0}(t)<\overline{e}_{i0}, \quad&\quad i\in\mathcal{V},
\end{aligned}
\right.
\end{eqnarray}
where $\underline{e}_{ii^+}\coloneqq d_{ii^+}^*-\underline{d}_{ii^+}$, $\overline{e}_{ii^+}\coloneqq \overline{d}_{ii^+}-d_{ii^+}^*$ and $\underline{e}_{i0}\coloneqq r-\underline{d}_{i0}$, $\overline{e}_{i0}\coloneqq \overline{d}_{i0}-r$.

\color{black}
Given the limited interaction range $\overline{d}_{ij}$ and collision-free operation threshold $\underline{d}_{ij}$ for each edge $(i,j)\in \overline{\mathcal{E}}$, we now provide conditions under which the desired formation pattern--characterized by the  radius $r$ and the prescribed inter-agent separation angle set $\boldsymbol{c}$--is considered feasible.

\begin{definition}[Feasible Formation]
	A desired formation pattern is said to be \textit{feasible} if the following two conditions are satisfied:
	\begin{itemize}
		\item The desired inter-agent and agent-target Euclidean distances satisfy the CM$\&$CF constraints, that is,
		$\underline{d}_{ii^+}<d_{ii^+}^*=2r\sin{\frac{c_{ii^+}}{2}}<\overline{d}_{ii^+}$ for $i\in \mathcal{V}\setminus\{N\}$ and $\underline{d}_{i0}<r<\overline{d}_{i0}$ for $i\in\mathcal{V}$. 
		
		\item The desired separation angle $c_{ii^+}$ between the $i$-th agent and the $(i+1)$-th agent lies within the open interval $(0^\circ, 180^\circ)$.
	\end{itemize}
\end{definition}

\begin{remark}
	Similar feasibility conditions can be found in \cite{mehdifar2020prescribed, fu2022distributed}. The first condition ensures that the desired distances meet the CM$\&$CF requirements, while the second condition guarantees that the  formation framework constructed in Section \ref{Construction of distance-based formation} retains its isostatic property (see Lemma \ref{isostatic_proof}), which is essential for the convergence analysis in Section \ref{V}.
\end{remark}
\color{black}

\subsection{Infinitesimally Rigidity ({\it IR}) Preservation Induced Error Constraints}\label{MIR}
Once the desired formation framework $\mathcal{F}^\ast=(\overline{\mathcal{G}},\overline{p}^*)$ is constructed to be infinitesimally rigid, then the small positive constant $\nu$ in Lemma \ref{Transformederrorconstraints} estimates the conservativeness for how far a framework $\mathcal{F}=(\overline{\mathcal{G}},\overline{p})$ can be from $\mathcal{F}^\ast$ such that {\it IR} of $\mathcal{F}$ is preserved. A sufficient condition for {\it IR} preservation in terms of formation errors is proposed in \cite{mehdifar2020prescribed}, and restated as follows,

\begin{lemma}[\cite{mehdifar2020prescribed}]\label{infinitesimalPreservation}
Consider two frameworks $\mathcal{F}^\ast=\left(\overline{\mathcal{G}},\overline{p}^\ast\right)$ and $\mathcal{F}=\left(\overline{\mathcal{G}},\overline{p}\right)$ sharing the same graph. If $\mathcal{F}^\ast$ is infinitesimally rigid and $\overline{\nu}$ is a sufficiently small positive constant satisfying: $\overline{\Psi}(\mathcal{F},\mathcal{F}^\ast) \coloneqq \sum_{(i,j)\in \overline{\mathcal{E}}}\max \{|\underline{e}_{ij}(0)|,|\overline{e}_{ij}(0)|\}\leq \overline{\nu}$, then securing $-{\underline{e}}_{ij}(t)<e_{ij}(t)<\overline{e}_{ij}(t)$ for all $t\geq 0$, where ${\underline{e}}_{ij}(t)$, $\overline{e}_{ij}(t)$ are both positive decreasing functions, guarantees that $\mathcal{F}$ is infinitesimally rigid for all time.
\end{lemma}
\begin{assumption}\label{initialIR}
The MAS and the target are initially positioned such that $\forall (i,j) \in \mathcal{\overline{E}}$, the initial formation error vector $e(0)=\mathrm{col}(e_{ij}(0)) \in \Omega_{IR}$, where $\Omega_{IR}=\{e(0) \in \mathbb{R}^{2N-1}\colon \overline{\Psi}(\mathcal{F},\mathcal{F}^\ast) \leq \overline{\nu} \}$. In addition,  $\underline{e}_{ij}(0)$ and $\overline{e}_{ij}(0)$ are chosen to be $|\underline{e}_{ij}(0)|=|\overline{e}_{ij}(0)|=|{e}_{ij}(0)|+\mu$ with $\mu$ also being a sufficiently small positive constant. 
\end{assumption}
\begin{remark}
Assumption \ref{initialIR} implies that the initial framework of the MAS is infinitesimally rigid around the target. Furthermore, the initial values for the two positive decreasing functions $\underline{e}_{ij}(t)$ and $\overline{e}_{ij}(t)$ are chosen such that $-\underline{e}_{ij}(0)<e_{ij}(0)<\overline{e}_{ij}(0)$ is satisfied. It can be observed from Lemma \ref{infinitesimalPreservation} that by maintaining 
\begin{equation}
	-\underline{e}_{ij}(t)<e_{ij}(t)<\overline{e}_{ij}(t), \forall t \geq 0,
\end{equation}
where $\underline{e}_{ij}(0)=\overline{e}_{ij}(0)=|{e}_{ij}(0)|+\mu$, 
the {\it IR} of the initial framework can be guaranteed. \textcolor{black}{The parameter  $\mu$ ensures  the infinitesimal rigidity of the actual formation framework. Decreasing  $\mu$ can increase the robustness against  formation distortions,  such as  flip and flex ambiguities. Increasing it reduces the conservativeness  in choosing the bounds $-\underline{e}_{ij}(0)$ and $\overline{e}_{ij}(0)$ for formation error $e_{ij}(t)$. }
\end{remark}

Now, considering the CM$\&$CF and {\it IR} preservation induced error constraints simultaneously, one can determine the following allowable upper bound and lower bound of formation error,
\begin{equation}\label{Errorbounds}
-\underline{e}_{ij}^*<e_{ij}(t)<\overline{e}_{ij}^*, \quad  \forall (i,j)\in \mathcal{\overline{E}},
\end{equation}
where $\underline{e}_{ij}^*=\min\{{\underline{e}}_{ij},\left|e_{ij}\left(0\right)\right|+\mu\}$, and $ \overline{e}_{ij}^*=\min\{{\overline{e}}_{ij},\left|e_{ij}\left(0\right)\right|+\mu\}$.
The squared distance errors are defined by 
\begin{equation}\label{sqerror}\eta_{ij}(t)=d_{ij}^2(t)-{d_{ij}^\ast}^2, \end{equation}
which can be equivalently written as $\eta_{ij}(t)=e_{ij}(t)(e_{ij}(t)+2{d_{ij}^\ast})$. Note that the formation errors $e_{ij}(t)=d_{ij}(t)-d_{ij}^*>-d_{ij}^*$, then it can be concluded that $\eta_{ij}(t)$ increases monotonously with respect to $e_{ij}(t)$ on its domain $[-d_{ij}^*, \infty)$. Now the CM$\&$CF and {\it IR} constraints in terms of $\eta_{ij}(t)$ can be derived using the bounds of formation errors in (\ref{Errorbounds}),
\begin{equation}\label{SquaredErrorbounds}
-\underline{\eta}_{ij}<\eta_{ij}(t)<\overline{\eta}_{ij},  \quad \forall (i,j)\in \mathcal{\overline{E}},
\end{equation} 
where $\underline{\eta}_{ij}\coloneqq \underline{e}_{ij}^*(-\underline{e}_{ij}^*+2{d_{ij}^\ast}), \overline{\eta}_{ij}\coloneqq \overline{e}_{ij}^*(\overline{e}_{ij}^*+2{d_{ij}^\ast})$, $\forall (i,j)\in\mathcal{\overline{E}}$.

\subsection{Performance Constraints }\label{ppfConstraints}
The prescribed performance function in this work imposed on each edge in $\overline{\mathcal{G}}$ is given with the following exponential form,
\begin{equation}\label{perFunc}
\beta_{ij}(t)=\left(\beta_{ij}\left(0\right)-\beta_{ij}\left(\infty\right)\right)e^{-\gamma_{ij}t}+\beta_{ij}\left(\infty\right),
\end{equation}
where $\beta_{ij}\left(0\right), \beta_{ij}\left(\infty \right), \gamma_{ij}$ are positive parameters.
Under (\ref{perFunc}), the bounds for the squared distance errors developed in (\ref{SquaredErrorbounds}) become the following time-varying bounds,
\begin{equation}\label{TVSquaredErrorbounds}
-\underline{\eta}_{ij}(t)<\eta_{ij}(t)<\overline{\eta}_{ij}(t), \quad \forall (i,j)\in \mathcal{\overline{E}}.
\end{equation} 
where $\underline{\eta}_{ij}(t)\coloneqq \underline{\eta}_{ij}\frac{\beta_{ij}(t)}{\beta_{ij}(0)}$ and $\overline{\eta}_{ij}(t)\coloneqq\overline{\eta}_{ij}\frac{\beta_{ij}(t)}{\beta_{ij}(0)}$ are two positive decreasing functions.
One can observe that if (\ref{TVSquaredErrorbounds}) is guaranteed, then  the monotonicity of the two positive decreasing functions $\underline{\eta}_{ij}(t)$ and $\overline{\eta}_{ij}(t)$ implies that
\begin{equation*}
  -\underline{\eta}_{ij}(0)< -\underline{\eta}_{ij}(t)<\eta_{ij}(t)<\overline{\eta}_{ij}(t)<\overline{\eta}_{ij}(0),
\end{equation*}
for $\forall (i,j)\in \mathcal{\overline{E}}$ and thus 
(\ref{SquaredErrorbounds}) also holds. Furthermore, the parameters $\beta_{ij}\left(0\right), \beta_{ij}\left(\infty \right), \gamma_{ij}$ can be designed to adjust the system's maximum overshoots, steady-states and convergence speeds, respectively. To get rid of the reliance on the time-varying bounds in (\ref{TVSquaredErrorbounds}), we define $\xi_{ij}(t)=\frac{\eta_{ij}(t)}{\beta_{ij}\left(t\right)}$. If $\xi_{ij}(t)$ stays within the following constant bounds, 
\begin{equation}\label{Xiconstaints}
-{\underline{\xi}}_{ij}<\xi_{ij}(t)<{\overline{\xi}}_{ij}, \quad \forall (i,j)\in \mathcal{\overline{E}},
\end{equation}
where $\underline{\xi}_{ij}=\frac{\underline{\eta}_{ij}}{\beta_{ij}(0)}$ and $\overline{\xi}_{ij}=\frac{\overline{\eta}_{ij}}{\beta_{ij}(0)}$, then the MAS satisfies the CM$\&$CF and {\it IR} constraints around the target.

\begin{assumption}\label{initialconditions}
Initially, the MAS (\ref{dyna}) and the target (\ref{target}) are positioned such that 
each unicycle in the MAS satisfies the CM$\&$CF and {\it IR} constraints with its neighbors and the target, that is, $-\underline{e}_{ij}^*<e_{ij}(0)<\overline{e}_{ij}^*,  \forall (i,j)\in \mathcal{\overline{E}}$.
\end{assumption}
Now, the target enclosing control problem considered in this work can be formally stated as follows.
\begin{problem}\label{controObjectives}
Under Assumptions \ref{initialIR} and \ref{initialconditions}, given a moving target (\ref{target}) and a desired formation framework $\mathcal{F}^\ast=\left(\overline{\mathcal{G}},\ {\overline{p}}^\ast\right)$ constructed in Section \ref{Construction of distance-based formation}, design control laws $v_i$, $w_i$ such that the unicycle-type MAS (\ref{dyna}) can achieve the following control objectives.
\begin{itemize}
	\item[i.]Each unicycle does not violate CM$\&$CF and {\it IR} constraints at any time, that is, $ -\underline{e}_{ij}^*<e_{ij}(t)<\overline{e}_{ij}^*,  \forall (i,j)\in \mathcal{\overline{E}}$ for all $ t\geq 0$.
	\item[ii.]The formation errors converge to zero asymptotically, that is, $\lim\limits_{t \rightarrow \infty} e_{ij}(t)=0,\forall \left(i,j\right)\in \overline{\mathcal{E}}$.
	\item[iii.]The velocity of each unicycle converges to the target’s velocity asymptotically, that is, $\lim\limits_{t \rightarrow \infty} \dot{p}_i(t)-v_0(t)=0, \forall i\in \mathcal{V}$.
\end{itemize}
\end{problem}

\section{Formation Control Laws}\label{IV}
In this section, we begin by introducing the transformed formation error and deriving its dynamics. Next, the virtual control inputs are proposed based on the transformed error system. Finally, linear and angular velocity control laws are designed to solve Problem {\ref{controObjectives}}.

\subsection{Transformed Formation Errors}\label{sigma}
Inspired by \cite{jin2018adaptive}, we introduce the following transformed error function for each edge in $\overline{\mathcal{G}}$,
{\small\begin{equation}\label{distanceTransformedError}
	\sigma_{ij}(\xi_{ij}(t))=\frac{{\overline{\xi}}_{ij}{\underline{\xi}}_{ij}\xi_{ij}(t)}{\left({\overline{\xi}}_{ij}-\xi_{ij}(t)\right)\left({\underline{\xi}}_{ij}+\xi_{ij}(t)\right)}, \forall (i,j) \in \overline{\mathcal{E}},
\end{equation}}

\noindent where $\sigma_{ij}(\xi_{ij}(t)):(-{\underline{\xi}}_{ij},\overline{\xi}_{ij})\mapsto(-\infty,\infty)$ is monotonously increasing.
It can be seen that $\sigma_{ij}(\xi_{ij}(t))$ reaches zero if and only if the desired formation shape is achieved. Moreover, the boundedness of $\sigma_{ij}(\xi_{ij}(t))$ implies that the constraints (\ref{Xiconstaints}) imposed on $\xi_{ij}(t)$ are satisfied, or equivalently, the MAS satisfies the CM$\&$CF and {\it IR} constraints around the target. 
For clarity of notations, we will omit the time index for variables that are implicitly time-dependent in the subsequent analysis. 

Next, the relationship between the transformed error $\sigma_{ij}$ and the velocity of the MAS $\dot{p}$ will be derived. First, taking the time derivative of the squared distance error yields $\dot{\eta}_{ij}=2p_{ij}^T(\dot{p}_i-\dot{p}_j), \forall (i,j)\in\overline{\mathcal{E}}$. Define 
$\eta=[\eta_{12}, \eta_{23},\ldots,\eta_{N^-N}, \eta_{10},\eta_{20},\ldots, \eta_{N0}]^T$, then the dynamics of the squared distance errors can be rewritten as 
\begin{equation}\label{dynamicsEta}
\begin{aligned}
	\dot{\eta}=&2R\left(\overline{p}\right)\dot{\overline{p}}\\
	=&	2\left[\begin{matrix}M&-M\boldsymbol{1}_N\otimes I_2\end{matrix}\right]\left[\begin{matrix}\dot{p}\\\dot{p}_0\\\end{matrix}\right]\\
	=&2M\dot{p}-2M\boldsymbol{1}_N\otimes v_0,
\end{aligned}	
\end{equation}
where $\left[\begin{matrix}M&-M\boldsymbol{1}_N\otimes I_2\end{matrix}\right]$ is the decomposition of the rigidity matrix $R(\overline{p})$ with $M\in\mathbb{R}^{(2N-1)\times2N}$ corresponding to the first $2N$ columns of $R(\overline{p})$, and $-M\boldsymbol{1}_N\otimes I_2 \in \mathbb{R}^{(2N-1)\times2}$ being the remaining two columns of $R(\overline{p})$. This decomposition can be verified by using (\ref{RigidityMatrix}). To introduce some properties of the sub-matrix $M$ that will be used in the convergence analysis of the closed-loop system, we establish the following lemma. 
\begin{lemma}\label{postivedef}
If the rigidity matrix $R(\overline{p})$ in (\ref{RigidityMatrix}) has full row rank,  
then the decomposed matrix $M$ also has full row rank, and $MM^T$ is also positive definite.
\end{lemma}
\begin{IEEEproof}
Rewrite the rigidity matrix as $R(\overline{p})=
MQ$, where $Q=\left[\begin{matrix}I_{2N}&-\boldsymbol{1}_N\otimes I_2\end{matrix}\right] \in \mathbb{R}^{2N\times(2N+2)}$. Since the rigidity matrix has full row rank, then $\mathrm{rank}(R(\overline{p}))=2N-1$.
From matrix analysis \cite{horn2012matrix}, one has that $\mathrm{rank}(R(\overline{p})) \leq \min\{\mathrm{rank}(M),\mathrm{rank}(Q)\}$. It can be observed that the rank of constant matrix $Q$ is $2N$ and that $\mathrm{rank}(M)\leq 2N-1$. Then one has $2N-1\leq\min\{\mathrm{rank}(M),\mathrm{rank}(Q)\}=\mathrm{rank}(M)\leq 2N-1$. Therefore, $\mathrm{rank}(M)=2N-1$, which implies that $M$ also has full row rank. Furthermore, given that $\mathrm{rank}(M)=\mathrm{rank}(MM^T)$ and $M$ has full row rank, we can conclude that $MM^T$ has no zero eigenvalues. Now, one has $x^TMM^Tx>0$ for any non-zero vector $x \in \mathbb{R}^{2N-1}$. Therefore, matrix $MM^T$ is positive definite.
	\end{IEEEproof}	
	From the definition of $\xi_{ij}=\frac{\eta_{ij}}{\beta_{ij}(t)}$, one has $\dot{\xi}_{ij}=\frac{1}{\beta_{ij}(t)}(\dot{\eta}_{ij}-\dot{\beta}_{ij}(t)\frac{\eta_{ij}}{\beta_{ij}(t)})=\frac{1}{\beta_{ij}(t)}(\dot{\eta}_{ij}-\dot{\beta}_{ij}(t)\xi_{ij})$. Define compact vector $\xi=\mathrm{col}(\xi_{ij})$ and two diagonal matrices $\beta(t)=\mathrm{diag}(\beta_{ij}(t))$ and $\dot{\beta}(t)=\mathrm{diag}(\dot{\beta}_{ij}(t))$. 
	Then, we can obtain
	\begin{equation}\label{dynamicsXi}
		\begin{aligned}
			\dot{\xi}=&\beta^{-1}(t)(\dot{\eta}-\dot{\beta}(t)\xi)\\
			=&\beta^{-1}(t)(2M\dot{p}-2M\boldsymbol{1}_N\otimes v_0-\dot{\beta}(t)\xi),
		\end{aligned}
	\end{equation}
	where the second equality is derived by substituting (\ref{dynamicsEta}) into (\ref{dynamicsXi}).
	Now taking the time derivative of $\sigma_{ij}$ results in
	\begin{equation}\label{dynamicsSigmaij}
		{\dot{\sigma}}_{ij}=\zeta_{ij}\left(t\right)\left(\dot{\eta}_{ij}-\dot{\beta}_{ij}(t)\xi_{ij}\right),
	\end{equation}
	where function $\zeta_{ij}\left(t\right) \coloneqq \frac{{\overline{\xi}}_{ij}{\underline{\xi}}_{ij}\left({\overline{\xi}}_{ij}{\underline{\xi}}_{ij}+\xi_{ij}^2\right)}{\left({\overline{\xi}}_{ij}-\xi_{ij}\right)^2\left({\underline{\xi}}_{ij}+\xi_{ij}\right)^2 }\frac{1}{\beta_{ij}(t)}>0$. Rewriting (\ref{dynamicsSigmaij}) in its compact form and combining it with (\ref{dynamicsEta}) yield
	\begin{equation}\label{dynamicsSigma}
		\begin{aligned}
			{\dot{\sigma}}=&\zeta(t)\left(\dot{\eta}-\dot{\beta}(t)\xi\right)\\
			=&\zeta(t)\left(2M\dot{p}-2M\boldsymbol{1}_N\otimes v_0-\dot{\beta}(t)\xi\right),
		\end{aligned}
	\end{equation}
	where $\sigma=\mathrm{col}(\sigma_{ij})\in \mathbb{R}^{2N-1}$ defines a column vector and  $\zeta(t)= \mathrm{diag}(\zeta_{ij}(t)) \in\mathbb{R}^{(2N-1)\times(2N-1)}$ defines a positive diagonal matrix. 
	
	\color{black}
	\subsection{Virtual Control Input}\label{desiredV}
	To design control laws suitable for unicycle-type agents, we first define a virtual control input $u_i$ that represents the desired velocity vector for agent $i$, assuming single-integrator dynamics. This virtual control input is established based on the transformed formation errors $\sigma$ developed in Section \ref{sigma} as follows,
	\color{black}	\begin{equation}\label{desiredv}
		u_i=\begin{bmatrix}
			u_{ix}\\u_{iy}
		\end{bmatrix}=-\sum\limits_{(i,j)\in{\mathcal{E}}}p_{ij}\zeta_{ij}(t)k_{ij}\sigma_{ij}-p_{i0}\zeta_{i0}(t)k_{i0}\sigma_{i0}+v_0,
	\end{equation}
	where $k_{ij}>0$ is the control gain. Moreover, the desired heading angle is calculated as 
	\begin{equation}\label{desiredA}
		\theta_{id}=
		\begin{cases}
			\atantwo(u_{iy},u_{ix}), \quad \mathrm{if} \quad u_i\neq 0;\\
			0, \quad\quad\quad\quad\quad\quad\quad \mathrm{if} \quad u_i= 0.
		\end{cases}
	\end{equation}
	Note that in (\ref{desiredv}), the first and second terms aim at regulating the inter-agent distances and agent-target distance, respectively, and the last term is for tracking the target. Correspondingly, define $u=[u_1^T,\ldots, u_N^T]^T\in \mathbb{R}^{2N}$, $K=\mathrm{diag}(k_{ij})\in \mathbb{R}^{(2N-1)\times(2N-1)}$, then the compact form of the virtual control input  becomes
	\begin{equation}\label{Desiredvc}
		u=-M^T\zeta(t)K\sigma+1_N\otimes v_0.
	\end{equation}
	\color{black}
 In the subsequent subsection,  the virtual control input $u_i$ is mapped onto the admissible motion space of the unicycle dynamics, yielding linear and angular velocity control laws for each agent.

	\color{black}
	\subsection{ Linear and Angular Velocity Control Laws}\label{twoControlLaws}
	Under the  virtual control input (\ref{desiredv}), the linear velocity control law for each unicycle is designed as follows,
	\begin{equation}\label{linearV}
		v_i=\frac{\|u_i\|}{\cos {e_{\theta_i}}},
	\end{equation}
	where 
	\begin{equation}\label{headingerror}
		e_{\theta_i}=\theta_i-\theta_{id} 
	\end{equation} denotes the heading error, i.e., the difference between the current heading angle and the desired heading angle in (\ref{desiredA}). It is noted that (\ref{linearV}) contains singularity points when $e_{\theta_i}=\pm\frac{\pi}{2}$. Therefore, the angular velocity control law needs to be designed to not only prevent the MAS from reaching these singularity points but also ensure that the heading errors converge to zero. To this end, the following angular constraints are imposed on the heading errors,
	\begin{equation}\label{AngularContraints}
		\left|{e_{\theta_i}}\right|<{\overline{e}_{\theta_i}},
	\end{equation}
	where ${\overline{e}_{\theta_i}}<\frac{\pi}{2}$ is the upper bound of $\left|{e_{\theta_i}}\right|$.
	Next, the barrier Lyapunov function methods will be utilized in the design of the angular velocity control law to handle angular constraints. Define the transformed angular error as 
	\begin{equation}\label{transformedANGLE}
		\sigma_{{e}_{\theta_i}}=\frac{{{\overline{e}^2_{\theta_i}}}e_{{\theta}_i}}{{{\overline{e}_{\theta_i}^2}-e_{{\theta}_i}^2}}.
	\end{equation}
	Similar to (\ref{distanceTransformedError}), function $\sigma_{{e}_{\theta_i}}$ is monotonously increasing with respect to  ${e}_{\theta_i}$ on $(-{\overline{e}_{\theta_i}},{\overline{e}_{\theta_i}})$ and its value 
	reaches zero if and only if the heading error approaches zero. It can also be observed that the boundedness of $\sigma_{{e}_{\theta_i}}$ implies the satisfaction of angular constraints (\ref{AngularContraints}), thus eliminating the singularities of (\ref{linearV}).
	Now, the following angular velocity control can be proposed,
	\begin{equation}\label{angularV}
		w_i=-k_{h1i}{{{\overline{e}^2_{\theta_i}}}}\sigma_{{e}_{\theta_i}}-k_{h2i}\left(\frac{{\overline{e}^2_{\theta_i}}-{e}_{\theta_i}^2}{{{\overline{e}^2_{\theta_i}}}}\right)^2 \lceil\sigma_{{e}_{\theta_i}} \rfloor^\frac{1}{2}+{\dot{\theta}}_{id},
	\end{equation}
	where $k_{h1i}, ~k_{h2i}$ are positive control gains. The calculations for ${\dot{\theta}}_{id}$ when $u_i\neq 0$ are given as follows,
	\begin{equation}
		{\dot{\theta}}_{id}=\frac{u_i^T \begin{pmatrix}
				0 & 1 \\ -1 & 0
			\end{pmatrix} \dot{u}_i}{\|u_i\|^2}=-\frac{u_i^T R_{\frac{\pi}{2}} \dot{u}_i}{\|u_i\|^2}.
	\end{equation}
	\begin{remark}
		Note that $\dot{u}_i$ can be obtained by taking the time derivatives of the terms in (\ref{desiredv}). In particular, the derivative of  $\zeta_{ij}$ has the following forms, 
		\begin{equation}\label{cal1}
			\dot\zeta_{ij}=\frac{\bar\zeta_{ij}\left(2p^T_{ij}(\dot{p}_i-\dot{p}_j)-\dot{\beta}_{ij}\xi_{ij}\right)-\zeta_{ij}\beta_{ij}\dot{\beta}_{ij}}{\beta_{ij}^2},
		\end{equation}
		where $\bar\zeta_{ij}\coloneqq\frac{2\overline{\xi}_{ij}\underline{\xi}_{ij}\xi_{ij}^3+6\overline{\xi}_{ij}^2\underline{\xi}^2_{ij}\xi_{ij}-2\overline{\xi}_{ij}^3\underline{\xi}_{ij}^2+2\overline{\xi}_{ij}^2\underline{\xi}_{ij}^3}{(\overline{\xi}_{ij}-\xi_{ij})^3(\underline{\xi}_{ij}+\xi_{ij})^3}$. Furthermore, the derivative of transformed error $\dot{\sigma}_{ij}$ and the agent's velocity $\dot{p}_i$ will be deduced in the following section.
	\end{remark}
\begin{remark}
	In the previous design procedure, it is necessary to assume that all agents can  sense the target’s position because of the following two critical requirements: 1) maintaining  network connectivity and ensuring collision avoidance among all agents and the target, and 2) the criteria for the distance-based formation to be {\it MIR}. To simplify our analysis, we also assume that each agent can directly access the target’s velocity through communication as specified in (\ref{desiredv}). However,  it is worth noting that this assumption can be relaxed to allow only a subset of agents to have knowledge of the target’s velocity by employing consensus-based velocity observers, as illustrated in \cite{Milad2020flocking}. Due to space limits, we did not include this aspect in the current study. 
\end{remark}
	\section{Analysis of the Closed-Loop System}\label{V}
	We start this section by first showing that the singularity problem can be avoided and that the heading angle errors converge to zero in fixed time under the angular velocity control law. Next, 
	the analysis of the closed-loop system, including the fulfillment of the CM$\&$CF and {\it IR} constraints and the convergence of the formation error, will be presented.

	\subsection{Fixed-time Convergence of Heading Angle Errors}\label{HeadingConvergence}
	\begin{theorem}\label{Theoremangle}
		Suppose that the initial heading angle error satisfies the constraints (\ref{AngularContraints}), that is, $\left|{e_{\theta_i}}(0)\right|<{\overline{e}_{\theta_i}}<\frac{\pi}{2}, \forall i \in \mathcal{V}$, then the heading angle errors (\ref{headingerror}) of MAS (\ref{dyna}) converge to zero in fixed time without violating the imposed angular constraints (\ref{AngularContraints}) under the  angular control law (\ref{angularV}), that is, $\lim\limits_{t \rightarrow T_{i}}{e_{\theta_i}}=0$, and $|{e_{\theta_i}}|<{\overline{e}_{\theta_i}}$ for all $t \geq 0$ and $i\in\mathcal{V}$.
	\end{theorem}
	\begin{IEEEproof}
	Consider a candidate Lyapunov function for each unicycle as 
	\begin{equation}\label{angleLya}
		V_{\sigma_{{e}_{\theta_i}}}=\frac{1}{2}\sigma_{{e}_{\theta_i}}^2.
	\end{equation}
	Taking the time derivative of (\ref{angleLya}) and combining \textcolor{black}{it} with (\ref{dyna}) and (\ref{angularV}) yield
		{
			\begin{equation}\label{derivangleLya}
				\begin{aligned}
					\dot{V}_{\sigma_{{e}_{\theta_i}}}=&{\sigma_{{e}_{\theta_i}}}\frac{d{\sigma_{{e}_{\theta_i}}}}{d{{{e}_{\theta_i}}}}{{\dot{e}_{\theta_i}}}=
					{\sigma_{{e}_{\theta_i}}}\frac{{\overline{e}^4_{\theta_i}}+{{\overline{e}^2_{\theta_i}}}e^2_{{\theta}_i}}{({{\overline{e}_{\theta_i}^2}-e_{{\theta}_i}^2})^2}(\dot{\theta}_i-\dot{\theta}_{id})\\
					=&\frac{{\overline{e}^4_{\theta_i}}+{{\overline{e}^2_{\theta_i}}}e^2_{{\theta}_i}}{({{\overline{e}_{\theta_i}^2}-e_{{\theta}_i}^2})^2}\bigg[-k_{h1i}{{{\overline{e}^2_{\theta_i}}}}\sigma^2_{{e}_{\theta_i}}\\
					&-k_{h2i}\left(\frac{{\overline{e}^2_{\theta_i}}-{e}_{\theta_i}^2}{{{\overline{e}^2_{\theta_i}}}}\right)^2 |\sigma_{{e}_{\theta_i}} |^\frac{3}{2}\bigg]\\
					=&-\frac{k_{h1i}{\overline{e}^6_{\theta_i}}}{({{\overline{e}_{\theta_i}^2}-e_{{\theta}_i}^2})^2}\sigma^2_{{e}_{\theta_i}}-k_{h1i}{\sigma^4_{{e}_{\theta_i}}}-k_{h2i}|{\sigma_{{e}_{\theta_i}}}|^{\frac{3}{2}}\\
					&-\frac{k_{h2i}{{e}^2_{\theta_i}}}{\overline{e}^2_{\theta_i}}|{\sigma_{{e}_{\theta_i}}}|^{\frac{3}{2}}.
				\end{aligned}	
		\end{equation}}
		
		\noindent First, we demonstrate that the imposed angular constraints are not violated by showing the boundedness of the transformed heading angular error $\sigma_{{e}_{\theta_i}}$. Recall from the initial condition that $|{e}_{\theta_i}(0)|<\overline{e}_{\theta_i}$, the barrier Lyapunov function is bounded at $t=0$, that is, there exists a constant $C>0$, such that ${V}_{\sigma_{{e}_{\theta_i}}}(0)<C$. Since all terms in (\ref{derivangleLya}) are non-positive, it can be concluded that ${V}_{\sigma_{{e}_{\theta_i}}}$ is non-increasing and ${V}_{\sigma_{{e}_{\theta_i}}}(t)\leq {V}_{\sigma_{{e}_{\theta_i}}}(0)$ for all $t>0$.
		Therefore, it can be concluded that for all $t>0$, the heading angle errors remain within the imposed constraints (\ref{AngularContraints}), thereby preventing the occurrence of the singularities in the linear velocity control law (\ref{linearV}). 
		
		Next, it can also be deduced from (\ref{derivangleLya}) that 
		\begin{equation}
			\begin{aligned}
				\dot{V}_{\sigma_{{e}_{\theta_i}}}	\leq & -k_{h1i}{\sigma^4_{{e}_{\theta_i}}}-k_{h2i}|{\sigma_{{e}_{\theta_i}}}|^{\frac{3}{2}}\\
				\leq & -4k_{h1i}V^2_{\sigma_{{e}_{\theta_i}}}-2^{\frac{3}{4}}k_{h2i}V^{\frac{3}{4}}_{\sigma_{{e}_{\theta_i}}}.		
			\end{aligned}
		\end{equation}
		In light of Lemma \ref{fixedtime}, the transformed angular errors will converge to zero in fixed time $T_{i}$, that is, $\lim\limits_{t \rightarrow T_{i}}{\sigma_{{e}_{\theta_i}}}(t)=0$, which implies that $\lim\limits_{t \rightarrow T_{i}}{{e}_{\theta_i}}(t)=0$ with the settling time for each agent being $T_{i}\leq T^{i}_{max}\coloneqq\frac{1}{4k_{h1i}}+\frac{4}{2^\frac{3}{4}k_{h2i}}$.
	\end{IEEEproof}
	
	\subsection{Analysis for the Boundedness and Convergence of Formation Errors}\label{FormationConvergence}
	Under the proposed linear velocity control law (\ref{linearV}), we can deduce from (\ref{dyna}) that 
	\begin{equation}\label{actualvelocity}
		\begin{aligned}
			\dot{p}_i&=\frac{\|u_i\|}{\cos {e_{\theta_i}}}\left[\begin{matrix}\cos(e_{\theta_i}+\theta_{id})\\ \sin(e_{\theta_i}+\theta_{id}) \end{matrix}\right]\\
			&=\left[\begin{matrix}\|u_i\|\cos\theta_{id}-\tan e_{\theta_i}\|u_i\|\sin\theta_{id} \\ \tan e_{\theta_i}\|u_i\|\cos\theta_{id}+\|u_i\|\sin\theta_{id}\end{matrix}\right]\\
			&=\begin{bmatrix}
				1 & -\tan e_{\theta_i}\\
				\tan e_{\theta_i} & 1
			\end{bmatrix}u_i\\
			&=(I_2+S_i)u_i,
		\end{aligned}
	\end{equation}
	where $S_i=\begin{bmatrix}
		0 & -\tan e_{\theta_i}\\
		\tan e_{\theta_i} & 0
	\end{bmatrix}$ is a skew matrix. Rewriting it in a compact form and substituting (\ref{Desiredvc}), one has 
	\begin{equation}\label{compactdotp}
		\dot{p}=(I_{2N}+S^*)(-M^T\zeta(t)K\sigma+1_N\otimes v_0),
	\end{equation}
	where $S^*=\mathrm{diag}(S_i)\in \mathbb{R}^{2N\times2N}$. It can be observed that $S^*$ is also a skew matrix and ${x^T}S^*x=0$, $\forall x \in \mathbb{R}^{2N}$.
	
	Now, we are prepared to present the second theorem, which states that under the proposed control laws, the control objectives in Problem \ref{controObjectives} can be achieved. 
	
	\begin{theorem}
		Under Assumptions \ref{initialIR} and \ref{initialconditions}, given a moving target (\ref{target}), suppose that the initial heading angle error satisfies the constraints (\ref{AngularContraints}), that is, $\left|{e_{\theta_i}}(0)\right|<{\overline{e}_{\theta_i}}<\frac{\pi}{2}, \forall i \in \mathcal{V}$, the control laws (\ref{desiredv}), (\ref{linearV}) and (\ref{angularV}) will enable the MAS (\ref{dyna}) to 
		\begin{itemize}
			\item[i.] always satisfy the CM$\&$CF and {\it IR} constraints among the agents and the target, that is, $ -\underline{e}_{ij}^*<e_{ij}(t)<\overline{e}_{ij}^*, \forall (i,j)\in \mathcal{\overline{E}}$ for all $ t\geq 0$;
			\item[ii.] converge asymptotically to the desired formation framework $\mathcal{F}^\ast=\left(\overline{\mathcal{G}},\ {\overline{p}}^\ast\right)$ around the target, that is, $\lim\limits_{t \rightarrow \infty} e_{ij}(t)=0,\forall \left(i,j\right)\in \overline{\mathcal{E}}$;
			\item[iii.] track the target with velocity converging to $v_0$, that is, $\lim\limits_{t \rightarrow \infty} \dot{p}_i=v_0$, $\forall i\in \mathcal{V}$. 
		\end{itemize}
	\end{theorem}
	\begin{IEEEproof}
		The proof follows a similar process as in \cite{mehdifar2020prescribed}. 
		First, define a Cartesian product of intervals $\mathcal{C}_{\xi}=\prod_{(i,j) \in \mathcal{\overline{E}}}(-\underline{\xi}_{ij}, \overline{\xi}_{ij})$. The closed-loop system in terms of $\xi$ can be obtained by substituting (\ref{compactdotp}) into (\ref{dynamicsXi}), that is,
		\begin{equation}\label{closed_loop}
			\begin{aligned} 		 	\dot{\xi}
				=&h(t,\xi)\\
				=&\beta^{-1}(t)\biggl(-2MM^T\zeta(t)K\sigma(\xi)-2MS^*M^T\zeta(t)K\sigma(\xi)\\
				&+2MS^*1_N\otimes v_0-\dot{\beta}(t)\xi\biggr),
			\end{aligned}
		\end{equation}
		where function $h(t,\xi): \mathbb{R}_{\geq0} \times \mathcal{C}_{\xi} \mapsto \mathcal{R}^{2N-1}$ is continuous on $t$ and locally Lipschitz on $\xi$ over the open set $\mathcal{C}_{\xi}$. Furthermore, it can be concluded from Assumptions \ref{initialIR} and \ref{initialconditions} that $\xi(0) \in \mathcal{C}_{\xi}$. 
		Then, the conditions of Lemma \ref{IVPsolutiuon} are satisfied. Applying Lemma \ref{IVPsolutiuon} guarantees the existence and uniqueness of a maximal solution $\xi(t) \in \mathcal{C}_{\xi}$ on a small interval near $t=0$, that is, 
		\begin{equation}\label{constraintsforaperiod}
			\xi_{ij}(t) \in (-\underline{\xi}_{ij}, \overline{\xi}_{ij}), ~~ \forall (i,j) \in \mathcal{\overline{E}}, ~\forall t\in [0,\delta),
		\end{equation} 
		which implies that the constraints imposed on the squared errors (\ref{SquaredErrorbounds}) are satisfied for $t\in [0,\delta)$. As a result, $-\underline{e}_{ij}^*<e_{ij}<\overline{e}_{ij}^*, \forall (i,j)\in\mathcal{\overline{E}}$ also holds for $t\in [0,\delta)$, which implies that the CM$\&$CF and {\it IR} constraints of framework $\mathcal{F}=(\overline{\mathcal{G}},\overline{p})$ are preserved. Moreover, since the $2N-1$ edges of the initial framework are maintained to be connected, thus $\mathcal{F}=(\overline{\mathcal{G}},\overline{p})$ is also minimally rigid for $t\in [0,\delta)$. Now we obtain from Lemma \ref{Rigidity properties} that the rigidity matrix $R(\overline{p})$ has full row rank and $R\left(\overline{p}\right)R\left(\overline{p}\right)^T$ is positive definite for $t\in [0,\delta)$. In light of the decomposition step of the rigidity matrix in (\ref{dynamicsEta}) and Lemma \ref{postivedef}, matrix $M$ also has full row rank and $MM^T$ is positive definite. Define the minimum eigenvalue of $MM^T$ as $\lambda_{m}$.
		
		Next, by substituting previously derived compact form of MAS's velocity (\ref{compactdotp}) into the dynamics of $\sigma$ in (\ref{dynamicsSigma}),
		one can obtain the closed-loop system in terms of the transformed errors as follows,
		{ \begin{equation}\label{closedloop}
				\begin{aligned}			\dot{\sigma}=&\zeta(t)\biggl[2M(I_{2N}+S^*)(-M^T\zeta(t)K\sigma+1_N\otimes v_0)\\
					&-2M\boldsymbol{1}_N\otimes v_0-\dot{\beta}(t)\xi\biggl]\\
					=&\zeta(t)\biggl(-2MM^T\zeta(t)K\sigma-2MS^*M^T\zeta(t)K\sigma\\
					&+2MS^*1_N\otimes v_0-\dot{\beta}(t)\xi\biggl).
				\end{aligned}
		\end{equation} }
		
		Consider a candidate Lyapunov function for the transformed formation errors as $V_{\sigma}=\frac{1}{2}\sigma^TK\sigma$. Taking its time derivative and substituting the dynamics of the transformed errors (\ref{closedloop}) yield
		{ \begin{equation}\label{FormationLya2}
				\begin{aligned}
					\dot{V}_{\sigma}=&\sigma^TK\zeta(t)\biggl(-2MM^T\zeta(t)K\sigma-2MS^*M^T\zeta(t)K\sigma\\
					&+2MS^*1_N\otimes v_0-\dot{\beta}(t)\xi\biggl)\\
					=&-2\sigma^TK\zeta(t)MM^T\zeta(t)K\sigma\\
					&+2\sigma^TK\zeta(t)MS^*1_N\otimes v_0-\sigma^TK\zeta(t)\dot{\beta}(t)\xi.
				\end{aligned}
		\end{equation}}
		
		\noindent The second equality holds due to the properties of the screw matrix $S^*$. From Young's Inequality, the second term in (\ref{FormationLya2}) satisfies that $2\sigma^TK\zeta(t)MS^*1_N\otimes v_0 \leq \sigma^TK\zeta(t)MM^T\zeta(t)K\sigma+(1_N\otimes v_0)^T(S^*)^TS^*1_N\otimes v_0 \leq \sigma^TK\zeta(t)MM^T\zeta(t)K\sigma +NC_{{e}_\theta} \bar{v}_0^2 $, where $C_{{e}_\theta}$ is defined as $C_{{e}_\theta}\coloneqq \max\limits_{i}\{\tan^2{e}_{\theta_i}\}$. 
		Now, (\ref{FormationLya2}) becomes 
		{\small \begin{equation}\label{FormationLya2_2}
				\begin{aligned}
					\dot{V}_{\sigma}
					\leq &-2\sigma^TK\zeta(t)MM^T\zeta(t)K\sigma+\sigma^TK\zeta(t)MM^T\zeta(t)K\sigma\\
					&+NC_{{e}_\theta} \bar{v}_0^2 -\sigma^TK\zeta(t)\dot{\beta}(t)\xi\\
					\leq & -\lambda_{m}\sigma^TK\zeta(t)\zeta(t)K\sigma+NC_{{e}_\theta} \bar{v}_0^2 -\sigma^TK\zeta(t)\dot{\beta}(t)\xi\\
					\overbrace{\leq}^{\textcircled{\raisebox{-.9pt} {3}}} & 
					-\left[\lambda_{m}-\frac{\epsilon^2}{2}\right]\sigma^TK\zeta^2(t)K\sigma+NC_{{e}_\theta} \bar{v}_0^2+\frac{\xi^T\dot{\beta}^2(t)\xi}{2\epsilon^2},
				\end{aligned}
		\end{equation}} 
		
		\noindent where $\frac{\epsilon^2}{2}<\lambda_{m}$ and the inequality ${\textcircled{\raisebox{-.9pt} {3}}}$ holds due to Young's Inequality. Denote the minimum eigenvalues of diagonal positive matrices $\zeta^2(t)$ and $K^2$ by $\lambda_{\zeta^2 }$ and $\lambda_{K^2}$, respectively, and define the maximum eigenvalue of matrix $\dot{\beta}^2(t)$ by $\lambda_{\dot{\beta}^2}$, then $\dot{V}_{\sigma}
		\leq -\left[\lambda_{m}-\frac{\epsilon^2}{2}\right]\lambda_{K^2}\lambda_{\zeta^2 }\|\sigma\|^2+NC_{{e}_\theta} \bar{v}_0^2+\frac{\lambda_{ \dot{\beta}^2}\|\xi\|^2}{2\epsilon^2}$, for $t\in[0,\delta)$. Moreover, it can be seen from (\ref{constraintsforaperiod}) that $\|\xi\|^2=\sum\limits_{(i,j)\in\overline{\mathcal{E}}}\xi_{ij}^2<\sum\limits_{(i,j)\in\overline{\mathcal{E}}}\max\{\underline{\xi}^2_{ij},\overline{\xi}^2_{ij}\}\coloneqq\kappa$. Let $\lambda\coloneqq \left[\lambda_{m}-\frac{\epsilon^2}{2}\right]\lambda_{ K^2}\lambda_{\zeta^2 }$, then $\dot{V}_{\sigma}
		\leq -\lambda\|\sigma\|^2+NC_{{e}_\theta} \bar{v}_0^2+\frac{\kappa}{2\epsilon^2}\lambda_{ \dot{\beta}^2}, \forall t \in [0,\delta)$. 
		It can be observed from Theorem \ref{Theoremangle} that $C_{{e}_\theta}<C_{\overline{e}_\theta}\coloneqq \max\limits_{i}\{\tan^2\overline{e}_{\theta_i}\}$, then $NC_{{e}_\theta} \bar{v}_0^2$ is bounded by $ NC_{\overline{e}_\theta} \bar{v}_0^2$. In addition, by definition of performance function (\ref{perFunc}), $\lambda_{\dot{\beta}^2}$ is also bounded. Now define a stacked vector $\psi=\mathrm{col}(C_{{e}_\theta} \bar{v}_0^2, \lambda_{\dot{\beta}^2}) \in \mathbb{R}^2$. Then using Lemma 11.3 of \cite{chen2015stabilization} ensures that there exists a $\mathcal{K}_{\infty}$ function $\alpha(\|\psi\|)$ such that $\dot{V}_{\sigma}
		\leq -\lambda\|\sigma\|^2+\alpha(\|\psi\|), \forall t \in [0,\delta)$. Now, we can derive from Theorem 2.7 of \cite{chen2015stabilization} that the closed-loop system is Input-to-State Stable (ISS) with respect to input $\psi$ for all $t\in [0,\delta)$. Then the boundedness of the input $\psi$ for all $t \geq 0$ implies that there exists an upper bound $\overline\sigma \in \mathbb{R}^+$ such that $\|\sigma\|\leq \overline\sigma<\infty$, therefore, $|\sigma_{ij}|\leq \overline{\sigma}$. By leveraging the monotonously increasing property of function $\sigma_{ij}(\xi_{ij}(t)) $in (\ref{distanceTransformedError}) on $(-\underline{\xi}_{ij}, \overline{\xi}_{ij})$ and the fact that $\lim\limits_{\xi_{ij}\rightarrow \overline{\xi}_{ij}}\sigma_{ij}(\xi_{ij})=\infty$, $\lim\limits_{\xi_{ij}\rightarrow -\underline{\xi}_{ij}}\sigma_{ij}(\xi_{ij})=-\infty$, this upper bound $\overline\sigma$ implies that there exists a new interval denoted by $(-\underline{\xi}^*_{ij}, \overline{\xi}^*_{ij})$ such that 
		\begin{equation}\label{newconstraints}
			-\underline{\xi}_{ij}<-\underline{\xi}^*_{ij}\leq\xi_{ij}(t)\leq\overline{\xi}^*_{ij}<\overline{\xi}_{ij},\quad \forall t \in [0,\delta),
		\end{equation}
		where $ \underline{\xi}^*_{ij} = -\min\{\arg\limits_{\xi_{ij}}|\sigma_{ij}(\xi_{ij})|=\overline\sigma\}$ and $\overline{\xi}^*_{ij} = \max\{\arg\limits_{\xi_{ij}}|\sigma_{ij}(\xi_{ij})|=\overline\sigma\}$. 
		
		Finally, we show by contradiction that the time interval $[0,\delta)$ can be extended to $[0,\infty)$. Based on (\ref{newconstraints}), we define another Cartesian product of intervals by $\mathcal{C}^*_{\xi}=\prod_{(i,j) \in \mathcal{\overline{E}}}[-\underline{\xi}^*_{ij}, \overline{\xi}^*_{ij}]$, which is a compact set. It can be observed that 
		\begin{equation}\label{contradictioninequal}
			\xi(t)\in\mathcal{C}^*_{\xi} \subset\mathcal{C}_{\xi}, \forall t \in [0,\delta).
		\end{equation} 
		Suppose that $\delta < \infty$, then from the definition of maximal solution $\xi(t) \in \mathcal{C}_{\xi}$ in (\ref{constraintsforaperiod}) and the continuity of the $\xi(t)$, there must exist a time instant $t^*\in [0,\delta)$ such that $\xi(t^*)\notin \mathcal{C}^*_{\xi}$, which raises a contradiction against (\ref{contradictioninequal}). Then, $\xi_{ij}(t) \in (-\underline{\xi}_{ij}, \overline{\xi}_{ij}), \forall (i,j) \in \mathcal{\overline{E}}$ holds for all $t\in [0,\infty)$. Therefore, the MAS satisfies the CM$\&$CF and {\it IR} constraints around the target during the whole evolution. 
		
		The analysis above also implies that (\ref{FormationLya2_2}) holds for all $t\geq 0$. Recall from Theorem \ref{Theoremangle} that the heading angle errors converge to zero in fixed time $T_{total}=\max\limits_{i}{T^{i}_{max}}, \forall i\in \mathcal{V}$, together with the fact that the derivative of the performance function $\dot{\beta}^2(t)$ converges to zero as $t\rightarrow \infty$, it can be concluded from (\ref{FormationLya2_2}) and input-to-state stability theory that $\lim\limits_{t\rightarrow\infty}{\sigma_{ij}}= 0, \forall (i,j) \in \mathcal{\overline{E}} $ and thus $\lim\limits_{t \rightarrow \infty}u_i=v_0$. 
		From the definition of $\sigma_{ij}$ in (\ref{distanceTransformedError}), one can conclude that $\lim\limits_{t\rightarrow\infty}{\xi_{ij}(t)}=\lim\limits_{t\rightarrow\infty}\frac{\eta_{ij}(t)}{\beta_{ij}(t)}=\lim\limits_{t\rightarrow\infty}\frac{d_{ij}^2(t)-{d_{ij}^\ast}^2}{\beta_{ij}(t)}=0$. Therefore, $\lim\limits_{t\rightarrow\infty}e_{ij}=0, \forall (i,j)\in\overline{\mathcal{E}}$. 
		Since $\lim\limits_{t \rightarrow \infty}u_i=v_0$ and $\lim\limits_{t \rightarrow T_{i}}e_{\theta_i}=0$, 
		the velocity of each unicycle (\ref{actualvelocity}) also converges to the target's velocity, that is, $\lim\limits_{t\rightarrow\infty}\dot{p}_i=\lim\limits_{t\rightarrow\infty}(I_2+S_i)u_i=v_0$.	
	\end{IEEEproof}
\color{black}
\begin{remark}
The proposed method differs from several recent studies addressing the cooperative target enclosing problem. 
Compared with \cite{fu2022distributed},  which focused on second-order MASs using artificial potential function methods, our approach addresses more complex  nonholonomic constraints and  achieves the desired formation task without encountering local minima issues. In \cite{ke2022fixed}, the number of formation vertices was required to match the number of controllable agents. In contrast, our framework incorporates the target as a formation vertex without requiring control over its motion, making it more suitable for real-world applications. Compared with \cite{lu2023cooperative}, which adopted  a feedback linearization method  for nonholonomic vehicles, our work employs a velocity decomposition technique such that the sensitivity to model uncertainties of the feedback linearization method can be avoided. Moreover, the results in \cite{lu2023cooperative} did not address connectivity maintenance among neighboring agents, which is explicitly handled in our proposed method.
\end{remark}
\color{black}
	\section{Simulation Example}\label{VI}
	Consider the target enclosing problem for five agents with unicycle dynamics, and the desired circular formation shape is described by radius $r=5~\mathrm{m}$ and the inter-agent separation angle vector $\boldsymbol{c}=[65^{\circ},75^{\circ},75^{\circ},80^{\circ},65^{\circ}]^\mathrm{T}$. 
	The desired distances between neighbors of the {\it MIR} formation framework $\mathcal{F}^*$ are computed as $d_{12}^*=2r\sin(\frac{65^{\circ}}{2})=5.373~\mathrm{m}$, $d_{23}^*=d_{34}^*=2r\sin(\frac{75^{\circ}}{2})=6.0876~\mathrm{m}$, $d_{45}^*=2r\sin(\frac{80^{\circ}}{2})=6.4279~\mathrm{m}$, and $d_{10}^*=d_{20}^*=d_{30}^*=d_{40}^*=d_{50}^*=r=5~\mathrm{m}$. The inter-agent interaction ranges and collision-free thresholds are given as $[\overline{d}_{12},\overline{d}_{23},\overline{d}_{34},\overline{d}_{45}]=[12,15,15,12]~\mathrm{m}$ and $[\underline{d}_{12},\underline{d}_{23},\underline{d}_{34},\underline{d}_{45}]=[0.6,0.6,1,1]~\mathrm{m}$, respectively. The agent-target interaction range and collision-free threshold of each agent are given as $\overline{d}_{i0}=15~\mathrm{m}$ and $\underline{d}_{i0}=0.8~\mathrm{m}$, respectively. 
	Moreover, the target is initially positioned at $p_0=[1.9,1.9]^T$ and moves with time-varying velocity $v_0(t)=[1,1.5\cos(0.1t)]^T$. The MAS is initially positioned at $p_1=[7.9,1.3]^T$, $p_2=[2.1,4.4]^T$, $p_3=[-6.5,4.3]^T$, $p_4=[-6.8,-5.4]^T$, and $p_5=[2.2,-7.0]^T$, with the corresponding initial heading angles set to $\theta_1=110^{\circ}$, $\theta_2=50^{\circ}$, $\theta_3=300^{\circ}$, $\theta_4=75^{\circ}$, and $\theta_5=110^{\circ}$. 
	The upper bound of the heading error is selected to be ${\overline{e}_{\theta_i}}=50^\circ$. It can be verified that the initial setups satisfy the requirements in Assumption \ref{initialconditions}. The sufficiently small positive constant $\mu$ in Assumption \ref{initialIR} is chosen to be $\mu=3$, and the performance function in 
	(\ref{perFunc}) is $\beta_{ij}(t)=(1-0.15)e^{-0.1t}+0.15$. \textcolor{black}{The control gains are chosen to be $k_{ij}=0.2$, $\forall (i,j)\in \overline{\mathcal{E}}$ and $k_{h1i}=k_{h2i}=0.5$, $\forall i\in\mathcal{V}$. In the current simulation, the prescribed performance function method and the CM$\&$CF constraints are incorporated to ensure that the transient response does not result in formation ambiguities. Accordingly, we set $\mu = 3$ to impose less restrictive requirements on the initial positions of the MAS. Additionally, the control gains  $k_{ij}$, $k_{h1i}$, and $ k_{h2i}$ are selected to appropriately balance the trade-off between achieving faster convergence and avoiding overly aggressive transient behavior. }

\begin{figure}[htp]
		\centering
		\includegraphics[width=\linewidth]{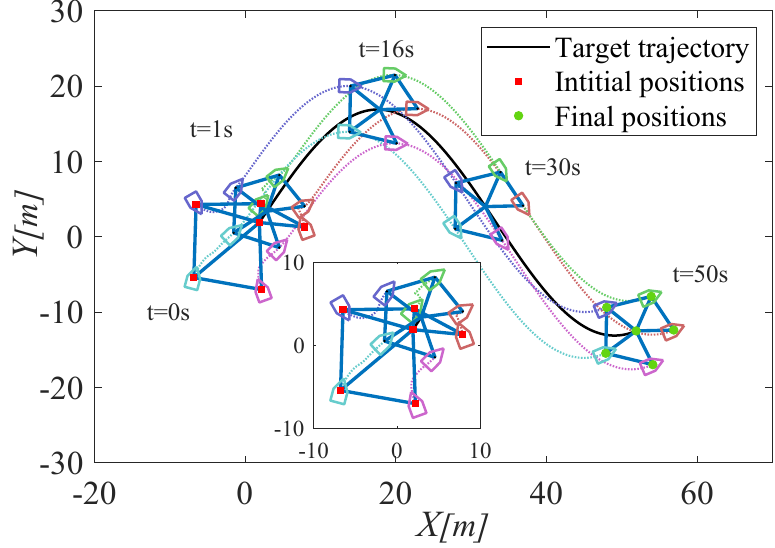}
		\captionsetup{justification=centering}
		\caption{The evolution of positions for the unicycle-type MAS and the target at $0$s, $1$s, $16$s, $30$s and $50$s.}
		\label{trajectories}
	\end{figure}

	The simulation is run for $50$ seconds and Fig. \ref{trajectories} shows the evolution of the positions for the unicycle-type MAS and the target on the mission space at time instants $0$s, $1$s, 	$16$s, $30$s and $50$s. The initial and final positions of the overall system are represented by the red squares and the green circles, respectively. The trajectory of the target  is indicated  by the black solid curve, while the agents' trajectories appear in dotted lines with varied colors for distinction. A zoomed inset of the figure, focusing on the interval from $t=0$s to $t=1$s,   is provided  at the mid-bottom to highlight the dynamic formation shape changes in the early stages.
	It can be observed that the desired formation shape around the target is finally achieved, with the heading angles of the unicycles precisely aligning with the target's velocity. 
	
\begin{figure}[htp]
		\centering
		\includegraphics[width=\linewidth]{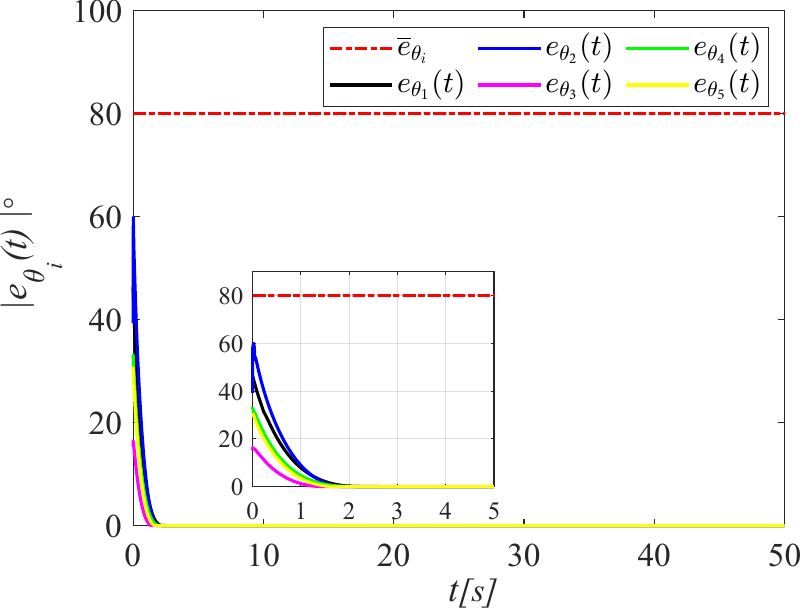}
		\vspace{-6pt}
		\captionsetup{justification=centering}
		\caption{The time responses of the heading angle errors. }
		\label{HeadingErrors}
		\vspace{-5pt}
	\end{figure}

	The time responses of the heading angle errors are shown in Fig. \ref{HeadingErrors}. The red dashed line represents the allowable maximum upper bound \textcolor{black}{${\overline{e}_{\theta_i}}=80^\circ$}. Fig. \ref{HeadingErrors} illustrates that under the proposed angular velocity control law $w_i$, the heading errors defined in (\ref{headingerror}) of the MAS converge to zero in fixed time. \textcolor{black}{A detailed view of the evolution from $t=0$ seconds to $t=5$ seconds is provided	in a zoomed-in segment of the figure.}
	
\begin{figure}[htp]
		\centering
		\begin{subfigure}{0.45\textwidth}
			\includegraphics[width=\linewidth]{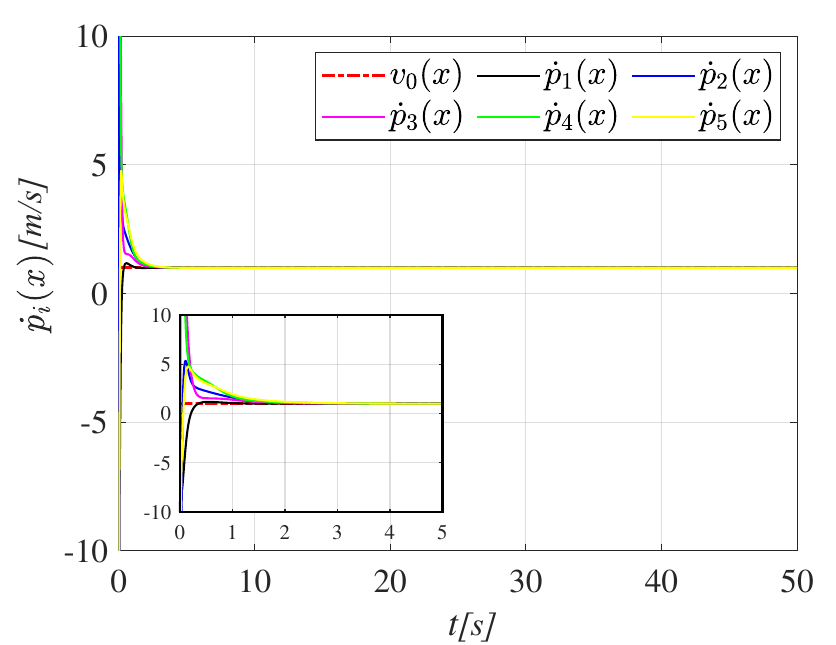}
			\caption{ Velocities along the X-axis.}
			\label{velocitiesX}
		\end{subfigure}
		\begin{subfigure}{0.45\textwidth}
			\includegraphics[width=\linewidth]{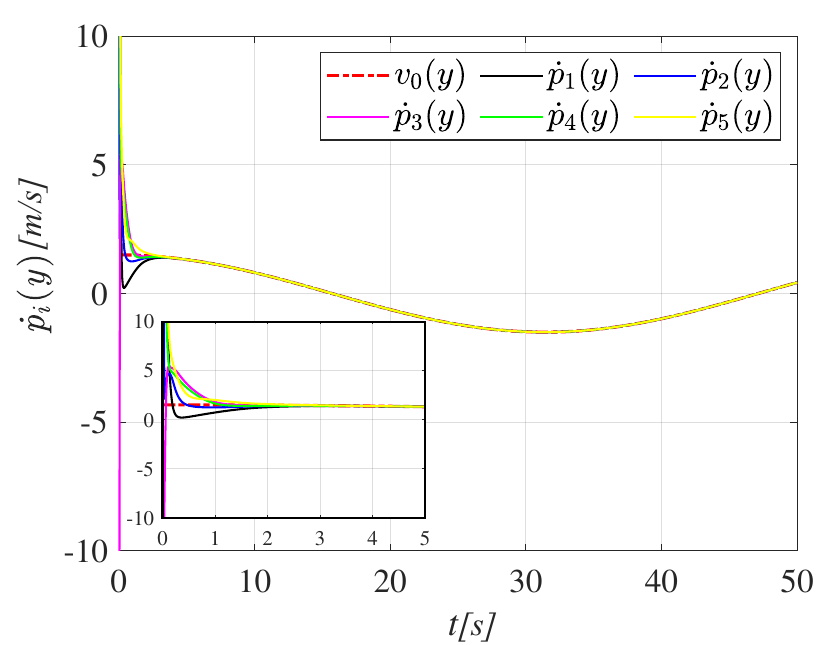}
			\caption{ Velocities along the Y-axis.}
			\label{velocitiesY}
		\end{subfigure}
		\caption{The time responses of the agents' velocities and the target's velocity $v_0$.}
		\label{velocities}
	\end{figure}

	The time responses of the agents' velocities and the target's velocity $v_0$ are shown in Fig. \ref{velocities}. Figs. \ref{velocitiesX}-\ref{velocitiesY} correspond to the evolution of the velocities along the X-axis and Y-axis, respectively. It can be observed that the 
	velocities of the MAS finally converge to the target's velocity $v_0$.
	
	Finally, to evaluate the satisfaction of CM$\&$CF and {\it IR} constraints of the proposed control laws, the two positive decreasing functions, $\overline{e}_{ij}(t)$ and $\underline{e}_{ij}(t)$ in Lemma \ref{infinitesimalPreservation}, can be calculated using the time-varying bounds of the squared distance error in (\ref{TVSquaredErrorbounds}) as follows, 
	{\small \begin{subequations}\label{errorBoundsFuncs}
			\begin{equation}
				\overline{e}_{ij}(t)=-d_{ij}^*+\sqrt{{d_{ij}^*}^2+\overline{\eta}_{ij}(t)},
			\end{equation}
			\begin{equation}
				\underline{e}_{ij}(t)=d_{ij}^*-\sqrt{{d_{ij}^*}^2-\underline{\eta}_{ij}(t)}.
			\end{equation}
	\end{subequations}}
	
\begin{figure*}[htbp]
	\centering
	\begin{subfigure}{0.3\textwidth}
		\includegraphics[width=\linewidth]{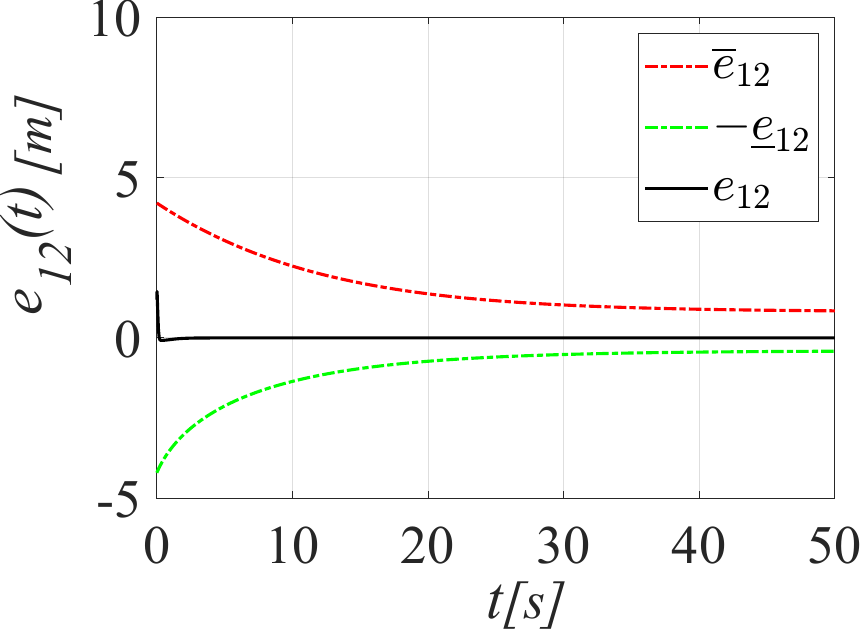}
		\caption{ $e_{12}(t)$. }
	\end{subfigure}
	\hfill
	\begin{subfigure}{0.3\textwidth}
		\includegraphics[width=\linewidth]{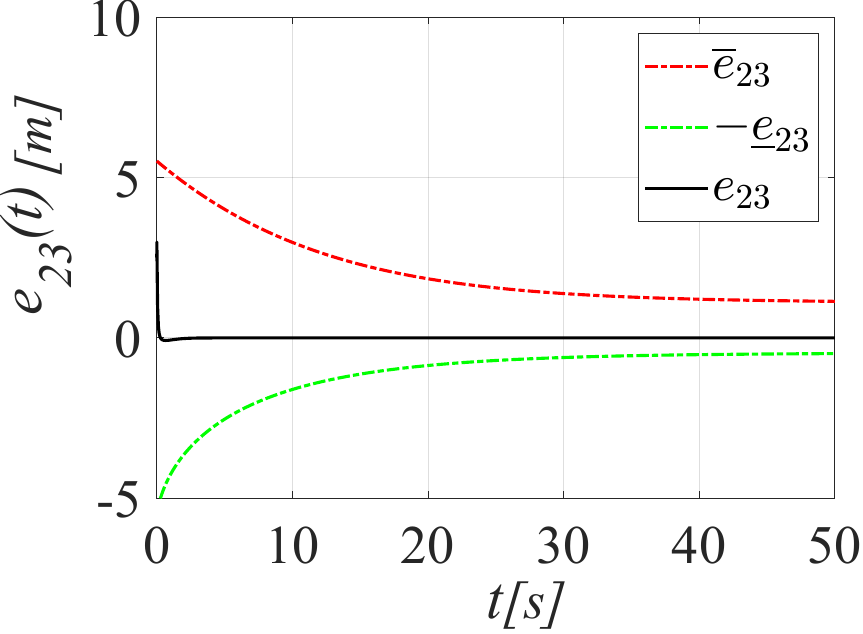}
		\caption{ $e_{23}(t)$.}
	\end{subfigure}
	\hfill
	\begin{subfigure}{0.3\textwidth}
		\includegraphics[width=\linewidth]{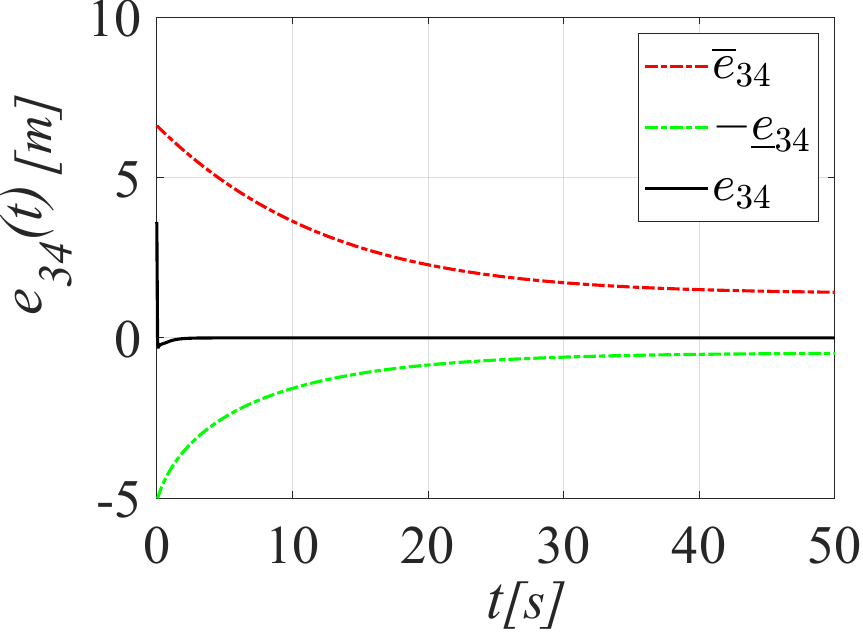}
		\caption{ $e_{34}(t)$.}
	\end{subfigure}
	
	\vspace{0.1cm} 
	
	\begin{subfigure}{0.3\textwidth}
		\includegraphics[width=\linewidth]{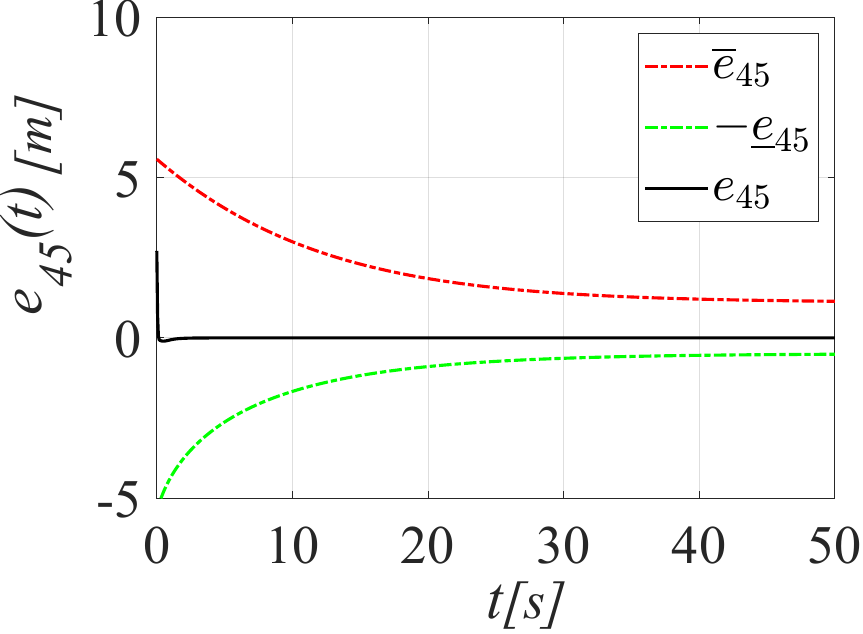}
		\caption{ $e_{45}(t)$.}
	\end{subfigure}
	\hfill
	\begin{subfigure}{0.3\textwidth}
		\includegraphics[width=\linewidth]{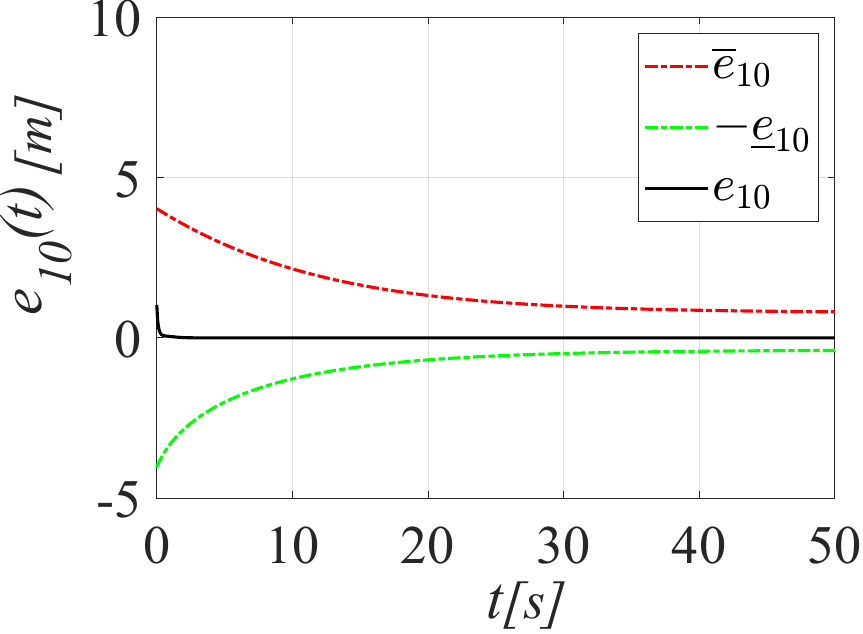}
		\caption{ $e_{10}(t)$.}
	\end{subfigure}
	\hfill
	\begin{subfigure}{0.3\textwidth}
		\includegraphics[width=\linewidth]{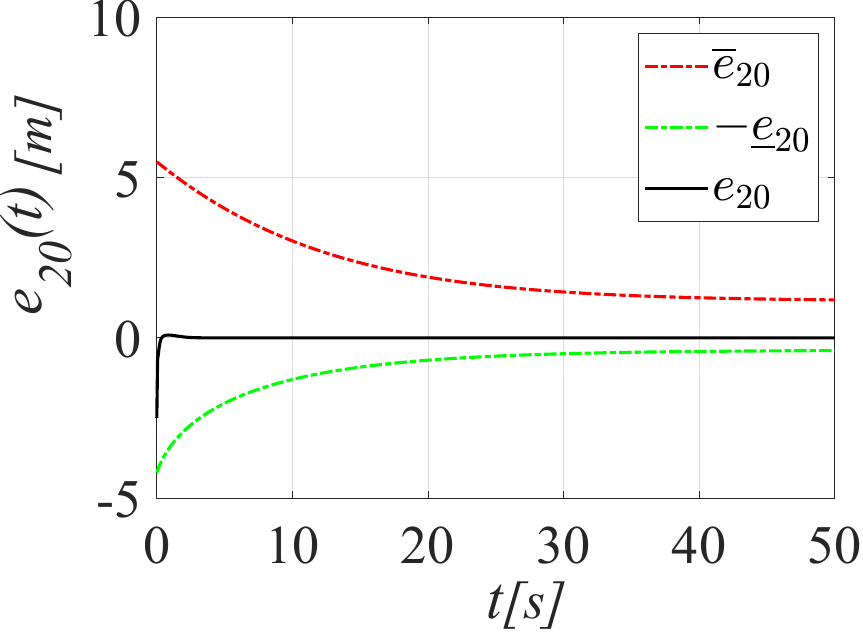}
		\caption{ $e_{20}(t)$.}
	\end{subfigure}
	
	\vspace{0.1cm}
	
	\begin{subfigure}{0.3\textwidth}
		\includegraphics[width=\linewidth]{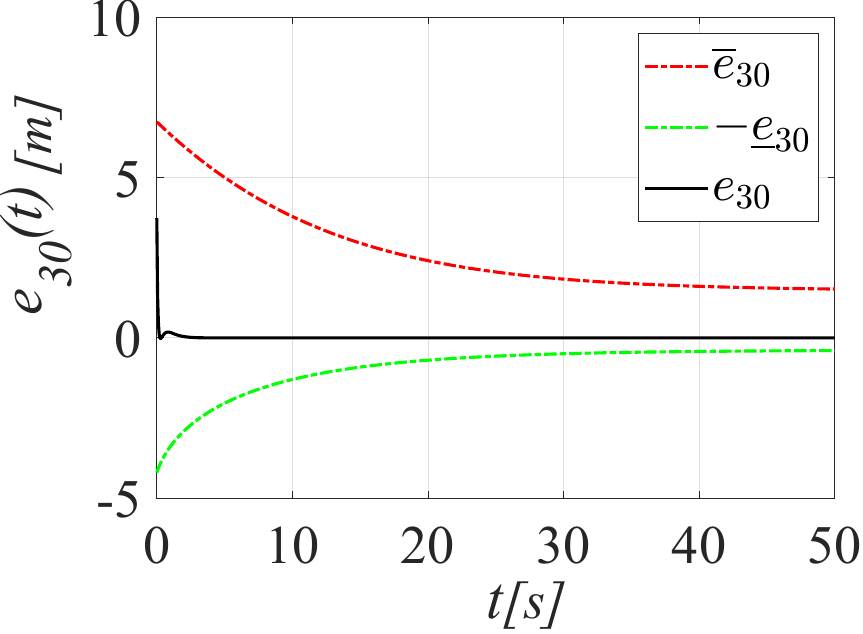}
		\caption{ $e_{30}(t)$.}
	\end{subfigure}
	\hfill
	\begin{subfigure}{0.3\textwidth}
		\includegraphics[width=\linewidth]{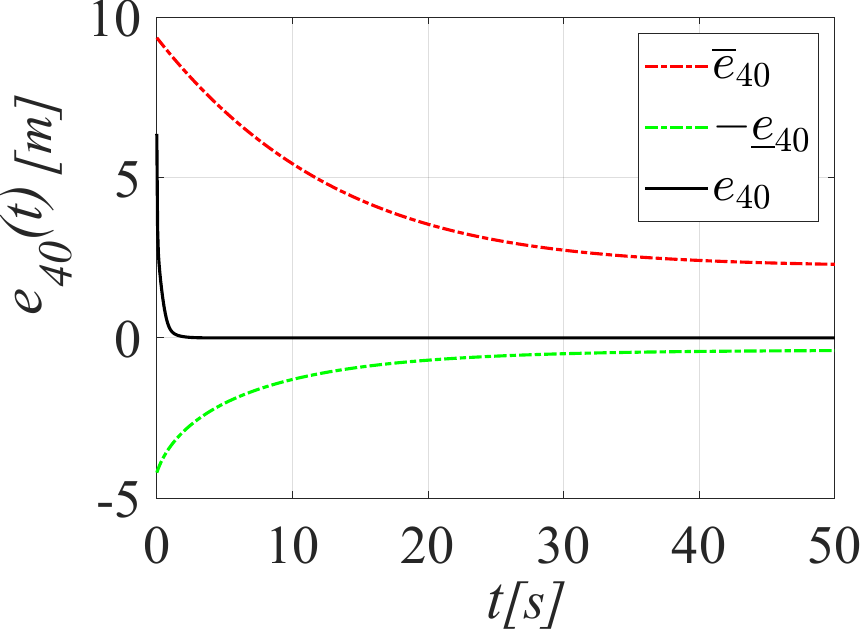}
		\caption{ $e_{40}(t)$.}
	\end{subfigure}
	\hfill
	\begin{subfigure}{0.3\textwidth}
		\includegraphics[width=\linewidth]{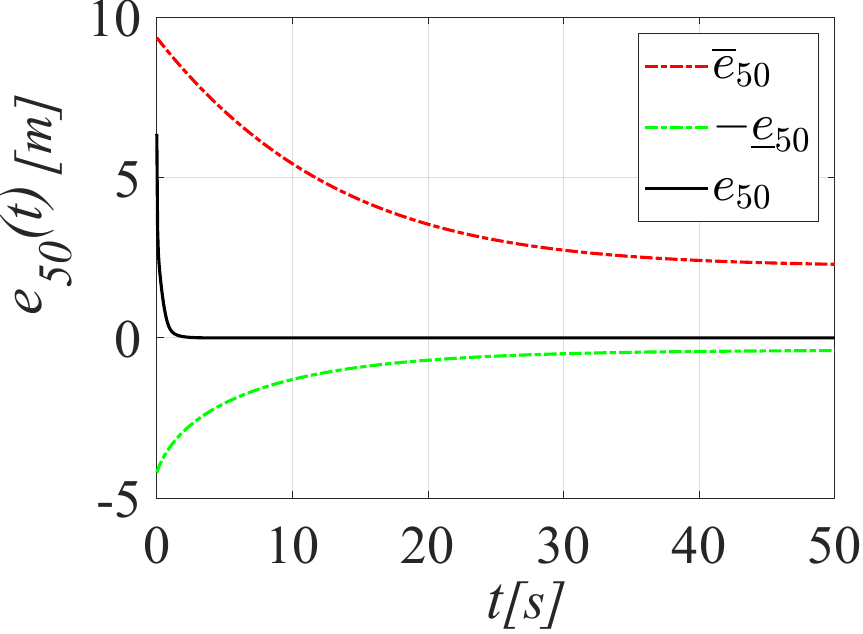}
		\caption{ $e_{50}(t)$.}
	\end{subfigure}
	\vspace{0.2cm}
	\caption{The time responses of the formation errors.}
	\label{constraints}
\end{figure*}

	\noindent Under the given simulation setups, it can be verified that $\overline{e}_{ij}(t)=-d_{ij}^*+\sqrt{{d_{ij}^*}^2+\overline{\xi}_{ij}\beta_{ij}(t)}$ and $\overline{e}_{ij}(t)=d_{ij}^*-\sqrt{{d_{ij}^*}^2-\underline{\xi}_{ij}\beta_{ij}(t)}$ hold. In Fig. \ref{constraints}, the time responses of the formation errors $e_{ij}(t)$, the time-varying upper bound function $\overline{e}_{ij}(t)$ and lower bound function $-\underline{e}_{ij}(t)$ for all edges in $\overline{\mathcal{G}}=\{\overline{\mathcal{V}}, \overline{\mathcal{E}}\}$ are plotted in black, red and green, respectively. The results depicted in Fig. \ref{constraints} indicate that formation errors converge to zero as time goes to infinity and that during the whole evolution, they always remain within the time-varying bound functions (\ref{errorBoundsFuncs}).

\section{Conclusion}\label{VII}
In this article, the moving target enclosing control problem for unicycle-type MASs has been investigated under the consideration of limited interaction ranges and collision avoidance. The desired target enclosing formation pattern has been carefully transformed into a distance-based formation framework, ensuring the resulting Gramian matrix of the rigidity matrix is positive definite. To address the challenges posed by distance constraints due to connectivity maintenance and collision avoidance, a transformed error system has been derived and the use of PPC techniques has further improved the convergence performance of the closed-loop system. Furthermore, the proposed angular control law, which leverages \textcolor{black}{the} barrier Lyapunov function method, has managed to eliminate the occurrences of the control singularity problem. The proposed control scheme guarantees that the MAS converges to the desired formation pattern while satisfying prescribed performance requirements, ensuring network connectivity, and avoiding collisions among neighboring agents and the target.

\textcolor{black}{In this work,  it is assumed that the interaction network is initially connected and network connectivity  is achieved by  each agent locally through maintaining connections of its initial edges, namely, the local connectivity maintenance approach. A recent study on maintaining global connectivity while executing a desired coordination task is proposed in \cite{ong2023nonsmooth}   and gives more flexibility on the dynamic changes of the underlying communication graph. In the future, the target enclosing problem  with global connectivity maintenance will be studied. 
Another promising direction is to consider scenarios where the target's velocity is not available to the MAS. In practical applications, measuring or obtaining this information in real time can be challenging. Therefore, the target enclosing control problem under distance constraints with unknown target motion will be an interesting topic.
A third direction arises from the limitations of the PPC method when velocity constraints are present.  In such cases, a new trade-off strategy needs to be developed to handle situations where the PPC-based controller exceeds velocity limits while still ensuring that safety-related distance constraints are satisfied. To this end, we also intend to incorporate a switching control protocol that tolerates temporary violations of the prescribed performance function but guarantees the satisfaction of safety constraints.}

{
\small
\bibliographystyle{IEEEtran}
\bibliography{IEEEabrv,MyCollection}
}

\vfill

\end{document}